%% file: jupring.tex
\documentclass{cplslarge}%
\usepackage[authoryear]{natbib}
\usepackage{color}
\usepackage{rotating}
\usepackage{dcolumn}  
\usepackage{amsfonts,amssymb,amsmath} 
\usepackage{makeidx}
\makeindex

\begin{document}

\newcommand{\m}{$\mu$m}     
\newcommand{\ms}{$\mu$m }     
\newcommand{\ares}{a_{\rm res}} 
\newcommand{\nres}{n_{\rm res}} 
\newcommand{\dd}{^\circ~{\rm d}^{-1}} 
\newcommand{\dpi}{\dot\varpi}
\newcommand{\dom}{\dot\Omega}
\newcommand{\dapi}{\Delta a_{\dpi}} 
\newcommand{\daom}{\Delta a_{\dom}} 
\newcommand{\dapat}{\Delta a_{p}} 
\newcommand{\dares}{\Delta a_{\rm res}} 
\newcommand{\gcm}{{\rm g~cm}^{-2}} 
\newcommand{\cmg}{{\rm cm}^2/{\rm g}} 
\newcommand{\et}{{\rm et al. }}

\renewcommand{\thechapter}{\arabic{chapter}}    
\setcounter{chapter}{5}

\setcounter{secnumdepth}{3}

\author[de Pater et al.]{I.~de~Pater, D.~P.~Hamilton, M.~R.~Showalter, H.~B.~Throop, and J.~A.~Burns}

\chapter{The Rings of Jupiter}

\section*{Book Chapter}

This is Chapter 6 in:  Planetary Ring Systems, Eds. M. S. Tiscareno and C. D. Murray. Cambridge University Press, Cambridge, UK. In Press (2017)

This version is free to view and download for personal use only. Not for re-distribution, re-sale or use in derivative works. Copyright: Cambridge University Press.
For more information on the book, see: www.cambridge.org/9781107113824.

\section{Introduction}

The jovian ring was discovered as the result of a concerted search by the
Voyager 1 cameras as the spacecraft passed Jupiter on March 4, 1979
\citep{smith79a}. Voyager's wide- and narrow-angle
cameras were targeted at the ring plane and exposed for 11 minutes. The
wide-angle field of view had complete coverage of the region, but was badly
overexposed by glare from Jupiter. However, by good fortune, the co-aligned
narrow-angle camera captured the tip of the ring. Subsequently, the Voyager 2
spacecraft performed a more detailed imaging sequence \citep{smith79b}. That
sequence included views looking back toward the Sun while passing through
Jupiter's shadow. In these views, what was hitherto a faint ring glowed
brightly, indicating that the ring is dominated by micron-sized dust grains,
which forward-scatter into a diffraction lobe just a few degrees wide. While
today we take it for granted that the majority of planetary rings are faint
and dusty, the jovian ring was the first ring found to be dominated by fine dust, and it
remains an archetype for similar dynamical systems.

Hints of the ring's existence had previously been obtained in 1974, when
Pioneer 11 detected ``drop-outs'' in the measurements of charged particles, 
attributed at the time to absorption by a potential ring around Jupiter 
\citep{fillius75, vanallen75}.
The rings have been subsequently imaged by three more spacecraft---Galileo,
Cassini, and New Horizons---as well as by the Hubble Space Telescope (HST)
and by several large, ground-based telescopes.

\begin{figure}
\figurebox{3.2in}{}{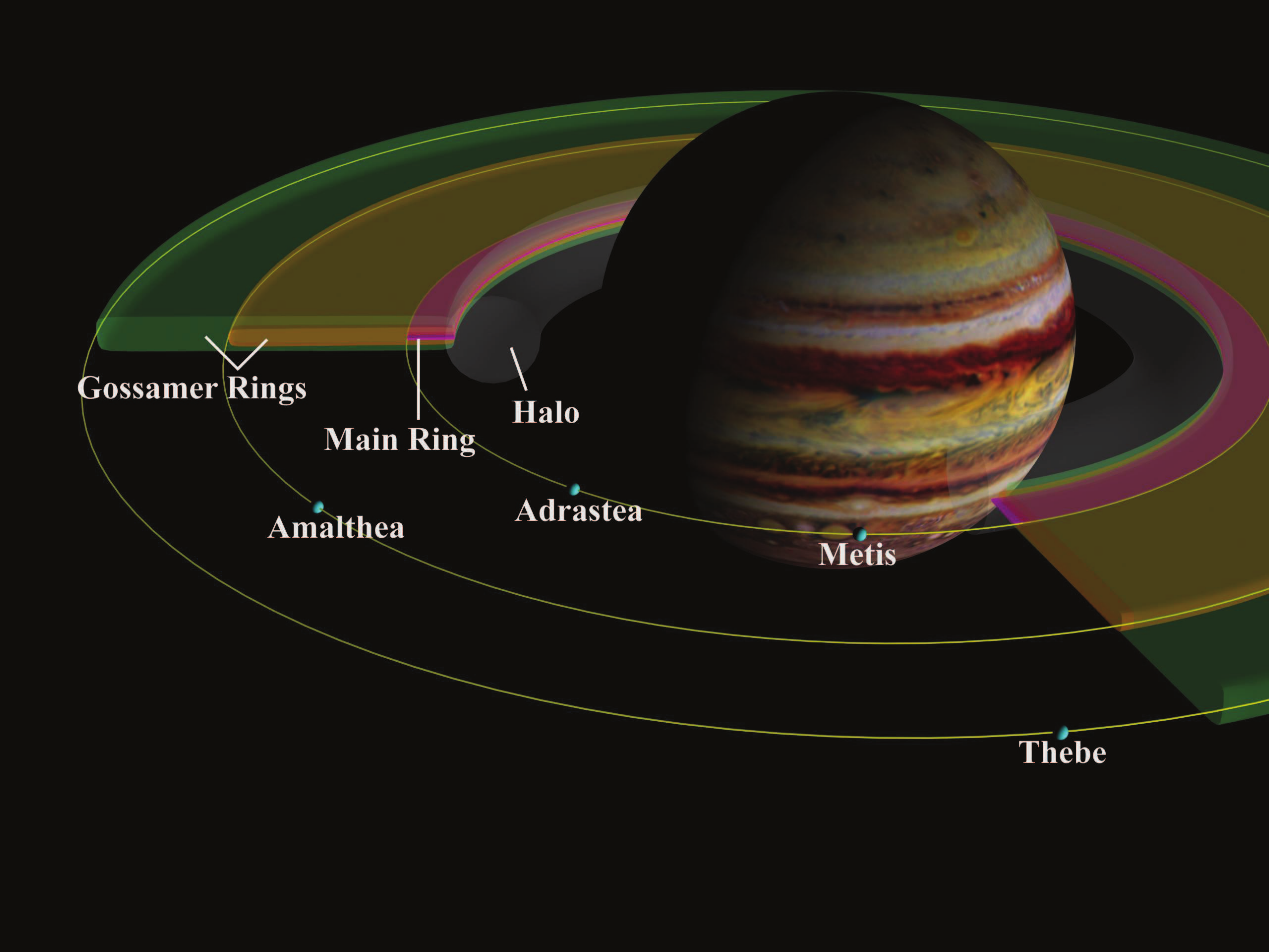}
\caption{Schematic of 
Jupiter’s ring system, and its relation to the jovian ring-moons. \citep{ockert-bell99}.
}
\label{fig:ringschematic}
\end{figure}

Jupiter's ring system consists of four principal components \citep{burns84,
showalter85, showalter87, ockert-bell99}: a) the `main ring' between $\sim$1.75
$R_\mathrm{J}$ and 1.81 $R_\mathrm{J}$ (where the Jovian radius $R_\mathrm{J} =$ 71,492 km) with optical depth $\tau
\approx \hbox{a few} \times 10^{-6}$ \footnote{Unless otherwise noted,
$\tau$ always refers to the normal optical thickness.}; b) the toroidal-shaped
`halo ring' interior to the main ring, which is a radially confined torus of faint
material with $\tau \approx 10^{-6}$; and c) the two faint `gossamer rings'
coinciding with the orbits of Amalthea and Thebe (both $\tau \sim 10^{-7}$;
Figure \ref{fig:ringschematic}, Table~\ref{tbl:tab1}) .

It is believed that the rings' optical depth is dominated by small
grains (radius $r < \lambda$, with $\lambda$ the observing
wavelength), which are produced by collisions or interactions with
larger parent bodies ( $r \geq \lambda$, and perhaps up to hundreds of
meters in size). These parent bodies have not been individually seen,
but their existence is indirectly inferred by the rings' existence, and
directly by measurements of the rings' phase curve and spectrum.

Since the jovian ring system is extremely faint, it is challenging to determine
its precise structure and the nature of the ring particles.  Information on the
rings' physical properties can be derived by obtaining data over a range of
phase angles $\alpha$ and wavelength $\lambda$; inversion of these spectra and
phase curves constrain the size and composition of both dust and
parent bodies. 

Since a major component of the rings is micron-sized dust, and the lifetimes of
such grains are brief due to the effects of solar radiation pressure, plasma
and electromagnetic perturbations ($\sim 10^3$--$10^5$ years for a 1~\ms grain,
\cite{burns04}), the rings must be young and continually replenished with new
material. \cite{burns99} suggested that impact ejecta from the inner
(ring-moon) satellites provide the sources for this ring material. They showed
that in particular the gossamer rings' morphology can be explained by ejecta
lost from Amalthea and Thebe on their inclined orbits, spiraling inwards due to Poytning
Robertson drag.

When rings are viewed edge-on (i.e., when the observer's elevation angle $B$ decreases to
$\sim 0^\circ$), relatively flat, optically thick rings fade as a result of
mutual shadowing and obscuration of ring particles. In contrast, the surface
brightness of optically thin rings brightens as 1/{\rm sin}$|B|$, since the
particles appear to be packed more closely as seen from the observer's
perspective. Hence, images both of the unlit and (nearly) edge-on rings provide
essential and complementary information to images taken under ``normal''
(back-scattered light) viewing conditions.  For Jupiter, Earth-based
telescopes, including HST, have imaged the rings
in backscattered light at phase angles $0^\circ$ -  $\sim 11^\circ$ and
wavelengths $0.5 \mu\rm{m} - 2.2 \mu\rm{m}$, while spacecraft have provided
data over the range $0^\circ$-$\sim 179^\circ$ and $0.5 \mu\rm{m} - 5
\mu\rm{m}$. {\it In situ} measurements of dust particles have further added to
the characterization of ring material.

Since an excellent review of Jupiter's rings has been provided by
\cite{burns04}, in the present review we focus on observations published since
that time, though we do provide some background to put the new
material in context.

\input{Table1.tex}

\section{The Main Ring and Halo} 

\subsection{Main Ring Morphology}

\begin{figure}
{\figurebox{3.2in}{}{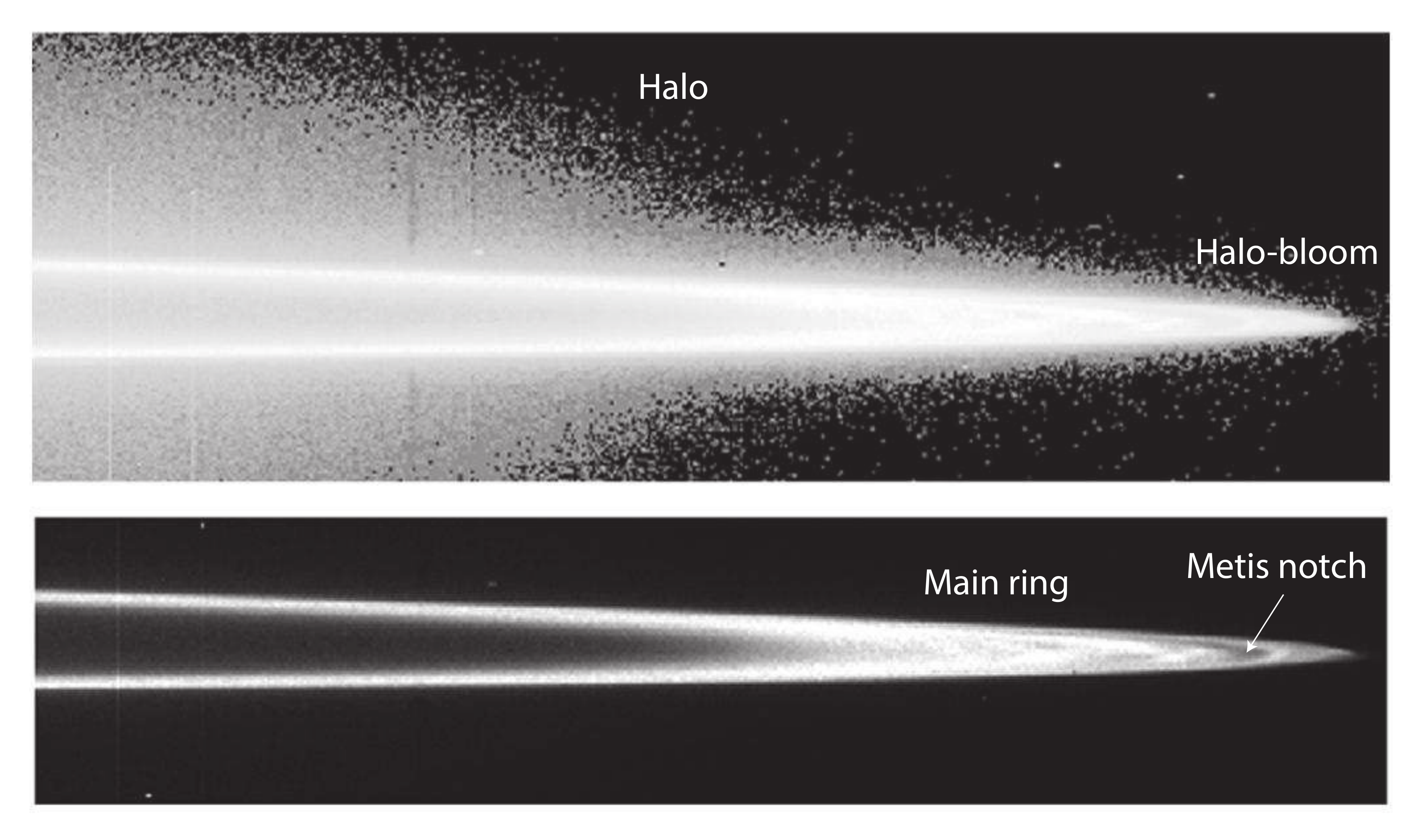}
\caption{Galileo view of Jupiter's main ring at a phase angle of 176$^\circ$.
The top image has been stretched to show both the main ring and the halo, while
the bottom image shows detailed patterns, such as the Metis gap, in the main
ring. (Adapted from \cite{ockert-bell99}.}

\label{fig:gal1}}
\end{figure}

\begin{figure}
{\figurebox{3.2in}{}{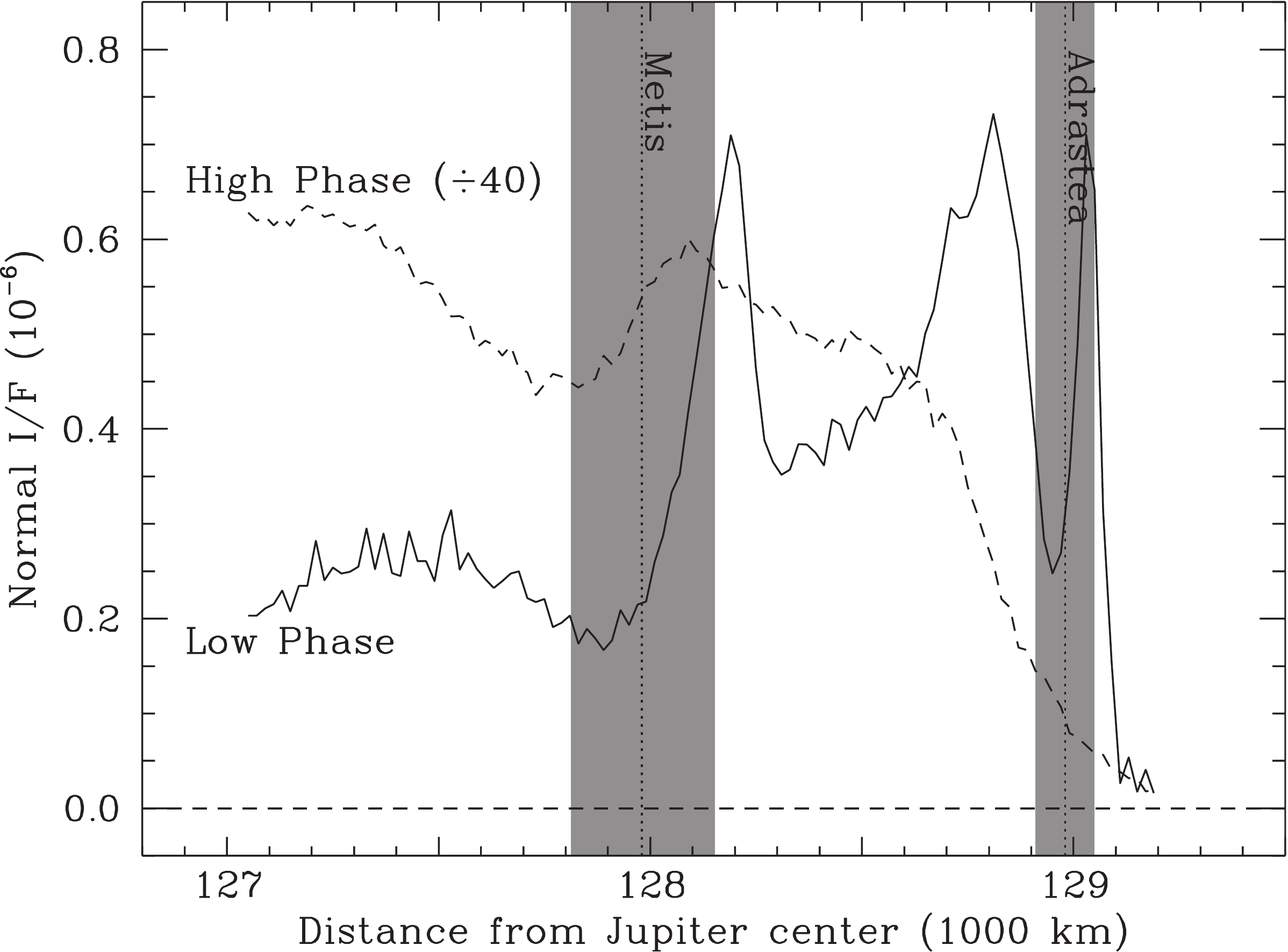}
\caption{Comparison of radial profiles of the annulus of the main ring in both
forward (high phase) and backscattered (low phase) light. In contrast to the
forward scattered light profile, the backscattered profile shows much structure
between the orbits of Metis and Adrastea, indicative of larger, parent bodies
within the system. Vertical lines locate the orbital semimajor axes of Adrastea
and Metis; the shaded vertical bands surrounding these orbital positions show
the sweeping zones of the satellites. (From \cite{burns04}, which was adapted
from \cite{showalter01}).}
\label{fig:gal2}}
\end{figure}

The main ring is the most prominent component of the jovian ring system,
especially when the rings are viewed in backscattered light, which is an
indication that there is a relatively larger fraction of macroscopic material
in this component of the rings compared to the fainter halo and gossamer rings.
The total radial extent of the main ring is $\sim 7000$~km, with a
$\sim$~800-km-wide bright annulus at its outer edge that has a sharp outer
boundary near 129,100 km (1.81 $R_\mathrm{J}$). The ring's normal optical depth $\tau
\approx \hbox{a few} \times 10^{-6}$. Spacecraft and ground-based images have
revealed detailed variations in the main ring's radial structure, such as shown
in the visible-light image, Fig.~\ref{fig:gal1}, taken with the Galileo
spacecraft in forward-scattered light.  Radial scans through the main ring
annulus in both forward and backscattered light are shown in
Fig.~\ref{fig:gal2}. Three bright radial bands are visible between the orbits of
Metis and Adrastea and just outside Adreasta in the low-phase profile, but
vanish in the corresponding high-phase images. These bands are thought to be
the parent-body population. 

Adrastea most likely clears a gap in the rings, while Metis controls the ring
annulus’ inner edge. No structure is visible interior to the orbit of Metis,
suggesting that there are no source bodies in this region; the region is
composed entirely of dust. 

\begin{figure}
{\figurebox{3.2in}{}{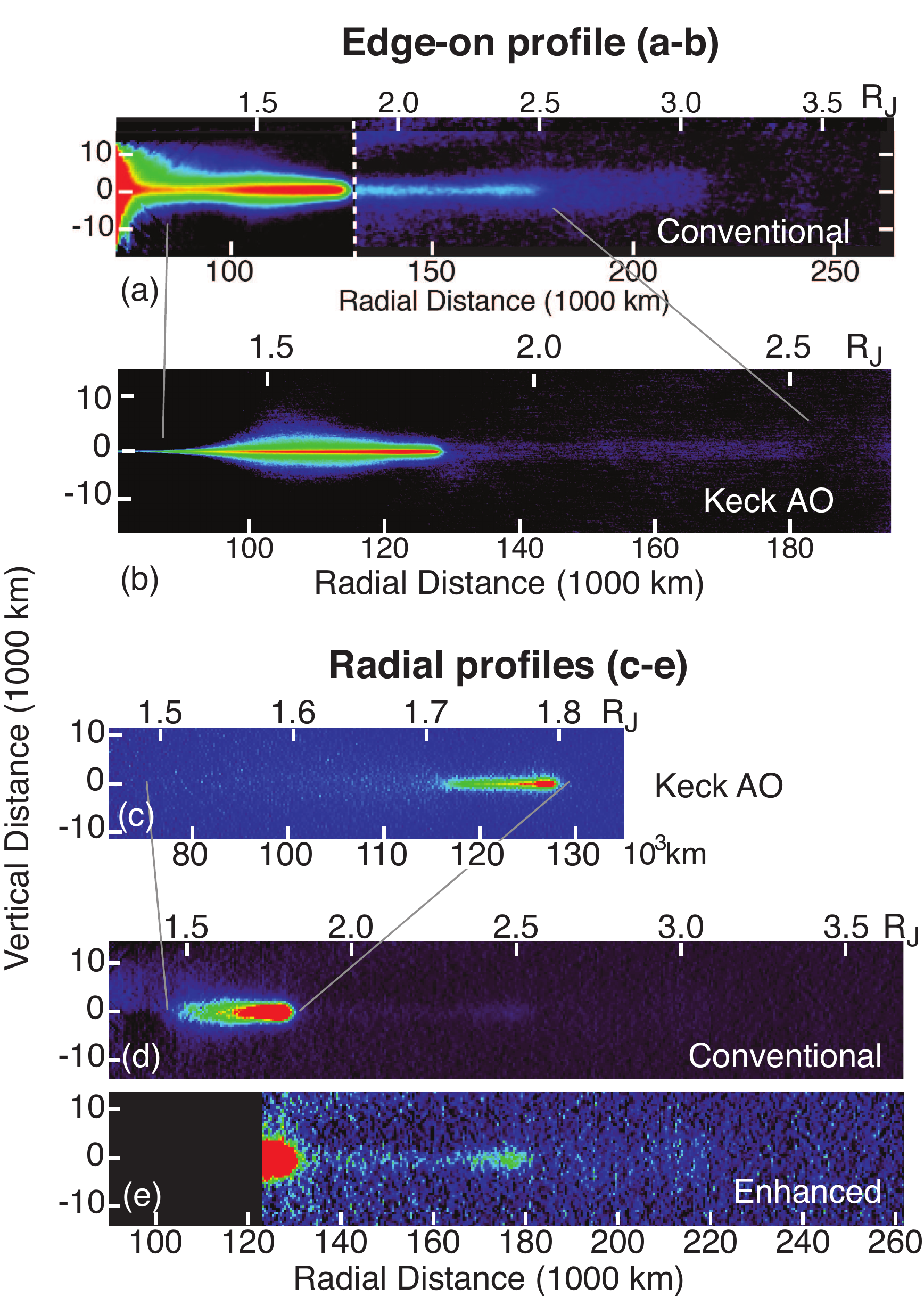}
\caption{Images of Jupiter's ring system taken with the Keck telescope. a)
Edge-on image of the rings at a wavelength of 2.2~\m; individual exposures were
taken on 19 Dec. 2002 and 22 Jan. 2003. The main ring is visible in red, the
halo in green, the inner gossamer (`Amalthea ring') in light-blue, and the
outer gossamer (`Thebe ring') in dark blue. Note that the vertical thickness of
the Thebe ring is much larger than that of the Amalthea ring. b) Edge-on image
of the rings taken with the adaptive optics (AO) system at a wavelength of 2.2~\ms on 26 Jan. 2003. 
The main ring is visible in red, the halo in green, and
the Amalthea ring in purple.  The decreased intensity in the latter ring at
the mid-plane is indicative of the same bright edges as seen in
Fig.~\ref{fig:ring-gos}a. c) Radial structure of the image in panel b, after
applying an ``onion''-peel technique, as described in the text. d) Radial
structure obtained from the edge-on image in panel a. e) Same image as in panel
d, but enhanced to show the Amalthea ring. The AO images have too low a
signal-to-noise to extract the radial profile of the Amalthea ring. (Adapted
from \cite{depater08}).}
\label{fig:keck1}}
\end{figure}

\begin{figure}
{\figurebox{3.2in}{}{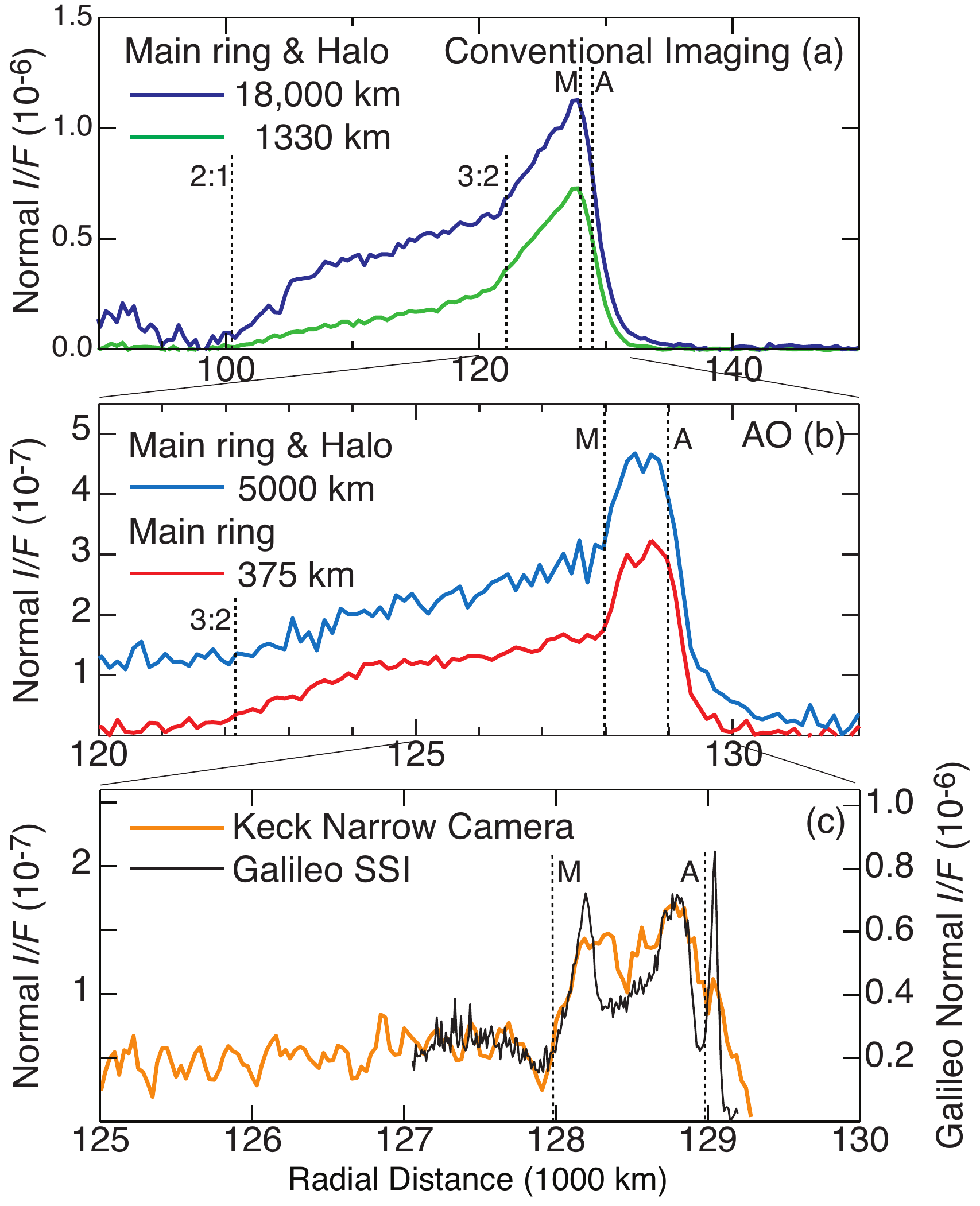}
  \caption{(a) Radial profiles through the main ring and halo,
    obtained by onion-peeling edge-on profiles obtained by integrating
    several rows in the image shown in Fig.~\ref{fig:keck1}a, (amounting to vertical widths of 18,000 km and 1330 km.  The
    orbits of Metis (M) and Adrastea (A), as well as the 3:2 and 2:1
    Lorentz resonances are indicated by dotted vertical lines. (b)
    Onion-peeled results from edge-on scans through the AO images in
    Fig.~\ref{fig:keck1}b. The upper profile in panel b shows the
    resulting radial profile for an edge-on scan that was vertically
    integrated over both the main ring and halo (vertical width of 5000 km). The lower profile
    shows the result for just the main ring (vertical width of 375 km). The orbits of Metis (M)
    and Adrastea (A), as well as the Lorentz 3:2 resonance, are
    indicated. (c) A high resolution radial profile through Jupiter's
    main ring, obtained by onion-peeling the edge-on profile from an
    image obtained with the highest resolution AO camera. The profile
    was smoothed radially over 0.03$^{\prime\prime}$. Superposed is the visible
    light Galileo profile at low phase angles from
    Fig.~\ref{fig:gal2}; its intensity scale is shown on the y-axis on
    the right. (From \cite{depater08}).}
\label{fig:keck2}}
\end{figure}

Figure~\ref{fig:keck1} shows several views of the jovian ring as observed with
the Keck telescope at a wavelength of 2.2~\ms in backscattered light, when the
rings were essentially edge-on. Because Jupiter's rings are optically thin, and
assuming them to be cylindrically symmetric, one can invert the edge-on images
by using an ``onion-peel'' deconvolution method to convert from integrated
line-of-sight images, to radial profiles (e.g., \citet{showalter87,
depater08}). Starting from the outermost pixel in each row, the normal
intensity of a narrow annulus that would produce the observed edge-on intensity
is calculated, and then subtracted from each interior pixel in that row. This
process is repeated until the entire image is inverted.
Figure~\ref{fig:keck1}c--d show images of the radial structure of the rings as
obtained via this process, i.e., each row in this image represents a radial
profile of the rings. To maximize the signal-to-noise in radial profiles,
several rows in the edge-on images were integrated and onion-peeling was
performed on these integrated edge-on profiles.  Figure~\ref{fig:keck2} shows
the resulting radial profiles.  The scans in panel a were obtained from a
conventional ``low-resolution'' image (Fig.~\ref{fig:keck1}a), and panels b-c
from images taken with adaptive optics techniques (panel b from
Fig.~\ref{fig:keck1}b, and panel c from an image with a higher spatial
resolution).  

Superposed on Fig.~\ref{fig:keck2}c is the Galileo visible-light
low-phase radial profile. The Keck 2.2~\ms and Galileo profiles are in
excellent agreement. The detailed structure in between and around the
two moons remains unexplained. A small satellite was proposed to cause
the low brightness between the satellites (Showalter \et 1987), but
the Keck data would have revealed moons 3--4 km in radius. New
Horizons performed an exhaustive search for smaller embedded moons,
with negative results down to a detection threshold of $\sim$0.5 km
\citep{showalter07}. This implies a sharp break in the size
distribution of small jovian moons, because the mass of the
next-largest ring-moon, which remains unseen, has less than $\sim$~10$^{-4}$ of
Adrastea's mass.

\begin{figure}
{\figurebox{3.2in}{}{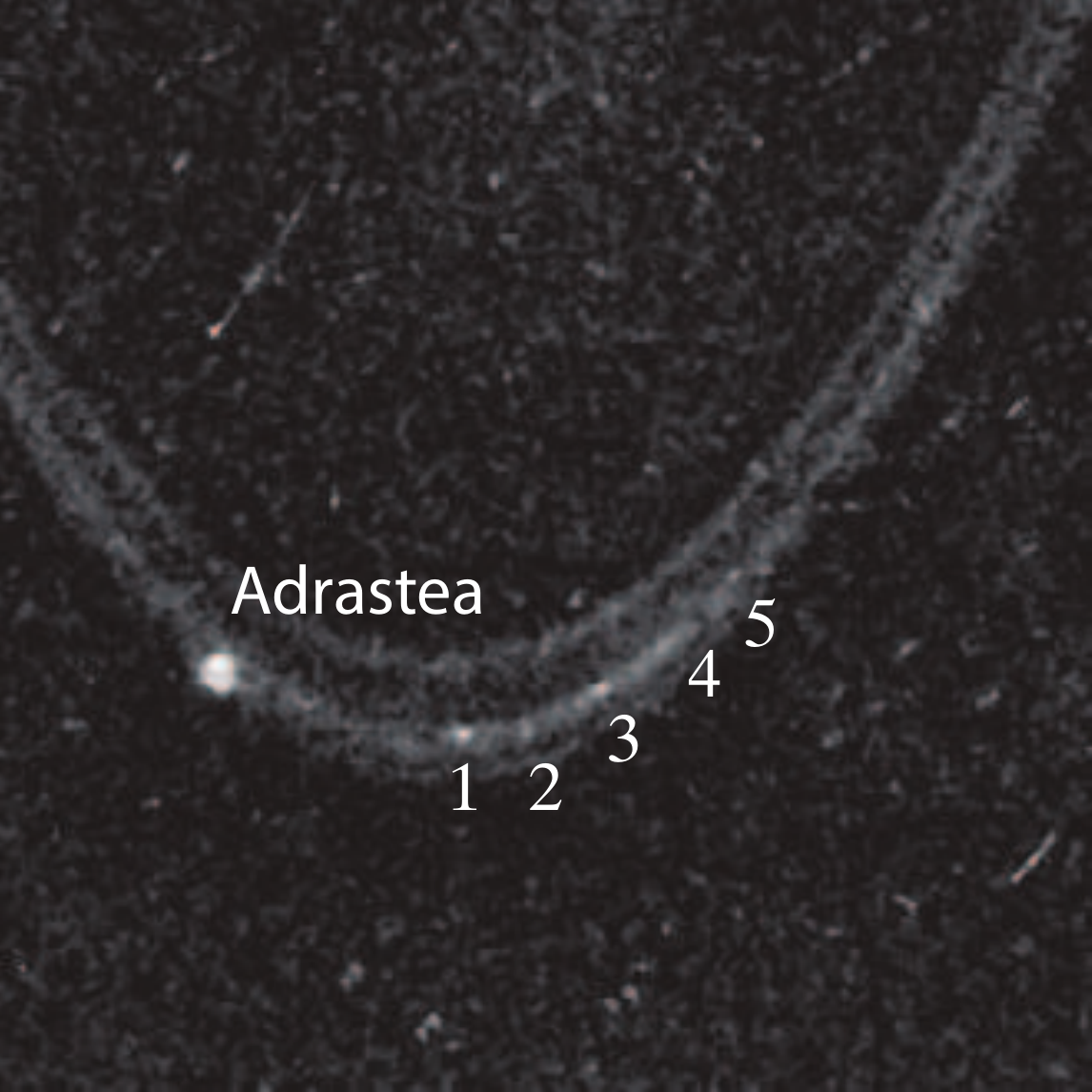}
\caption{Image obtained with the New Horizons spacecraft en route to Pluto. The
main ring annulus, including the central dip and the Adrastea gap, is clearly
visible, as is Adrastea and the presence of 5 small clumps (1-5). (From
\citep{showalter07}). 
}
\label{fig:NH}}
\end{figure}

The New Horizons images did, however, reveal several clumps embedded within
the rings. Because these
clumps are spread out in longitude over $0.1^\circ$--$0.3^\circ$, the bright
specks in Fig.~\ref{fig:NH} are not small moons, but rather distinct azimuthal
clouds of material. These concentrations were not seen in
forward-scattered light, so the clumps must be composed of larger-sized material.
If this material is similar to that of the moon Adrastea, with a geometric
albedo of 0.05, then the brightest clumps have an integrated surface area
equivalent to a moon 0.5 km in radius. Previously, Throop \et (2004) had found
evidence for a single, longer ($\sim$1000 km) clump in the Cassini images,
but it remains unclear if these are observations of the same phenomenon.

\subsection{Halo}

The halo ring is an optically thin torus inward of Metis. The diffuse ring is
visible in many images of the ring system, but no distinct features have been
found within the ring.  Interior to Metis, the ring's optical depth $\tau$
(Fig.~\ref{fig:keck2}) decreases steeply down to approximately the 3:2 Lorentz
resonance at 122,400 km.  Inside of this resonance, the ring's brightness
decreases more gradually, and seems to fade into the background inside the 2:1
Lorentz resonance at 101,200 km.  \cite{showalter01} showed that the halo's
brightness varies as a power-law with height, $z$, above the ring plane.
Closest to the ring plane, the brightness is proportional to $z^{-0.6}$, and
several 1000 km up it follows a power law $z^{-1.5}$.  Most of the halo's
material is thus concentrated within just a few hundred km of the ring plane.

The Voyager and Galileo forward-scattered profiles suggest that the material
interior to the main ring annulus is composed of micron-sized dust. This dust
is produced from the parent bodies, and transported inwards by
Poynting-Robertson drag. The particles within the halo probably had their
inclinations increased by interactions with the planet's magnetic field at
Jupiter's 3:2 Lorentz resonance. A second interaction with the 2:1 Lorentz
resonance can perturb particle orbits into Jupiter's atmosphere. Inside of the
2:1 resonance, the particle density is too low to be detectable from the ground.
Galileo images have revealed a faint extended halo closer to the planet, but
imaging becomes difficult because of the increase in telescopic stray light
moving closer to Jupiter's limb \citep{throop2016}.

Finally, exterior to the halo is a faint glow referred to as ``halo bloom,''
which was only noted in Galileo images
(Fig.~\ref{fig:gal1}) \citep{ockert-bell99}. This material has not been studied
further, but it may simply represent an outer, low-altitude component of the halo.

\subsection{Particle Size Distribution}

\begin{figure}
{\figurebox{3.2in}{}{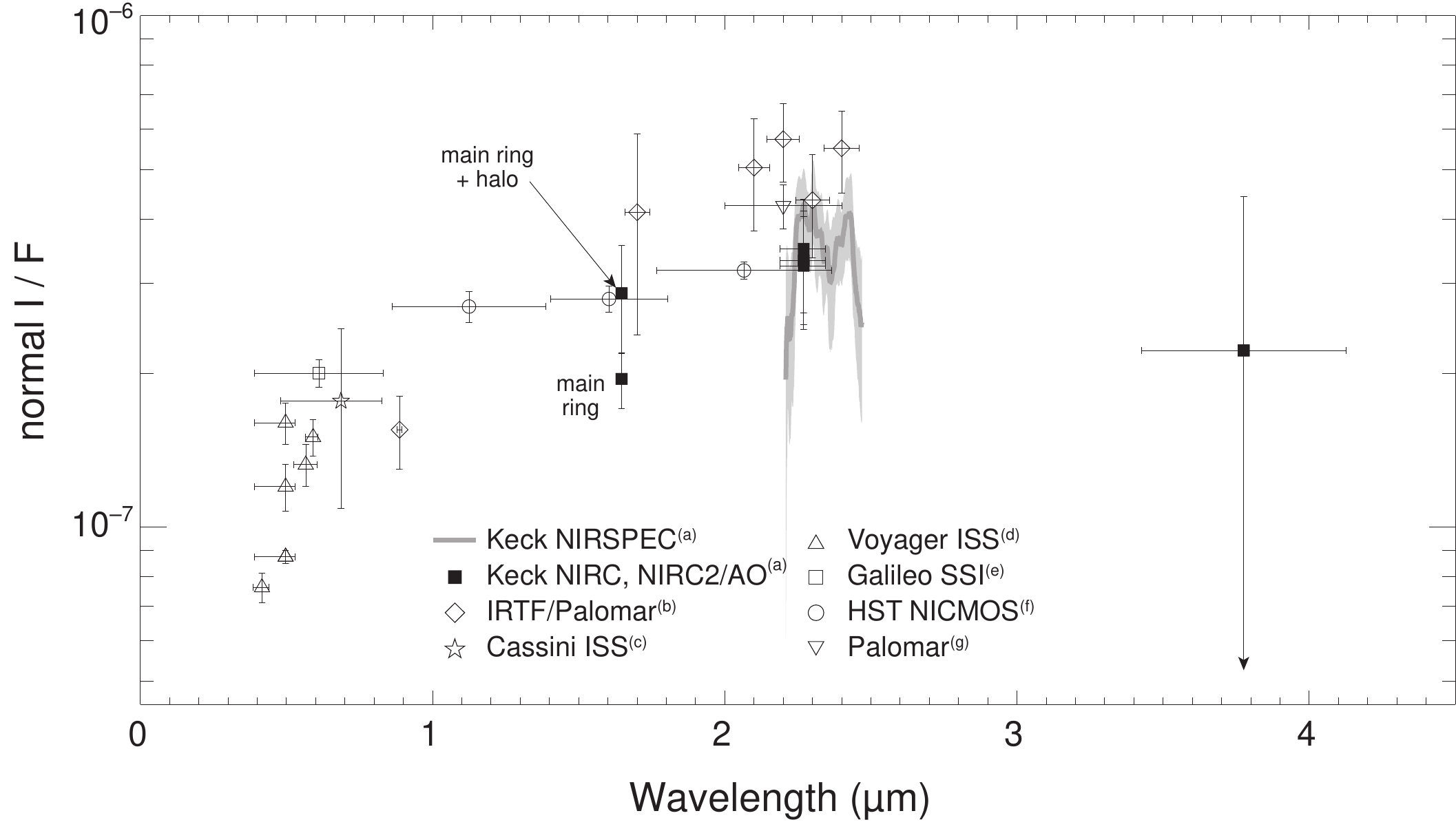}
\caption{Spectral measurements of the back-scattered main ring reflectivity,
after corrections for variations in the viewing geometry (by Throop \et 2004).
Horizontal bars indicate the wavelength range of each measurement. A Keck
NIRSPEC spectrum is shown as a dark grey curve, with a light grey error
envelope. The Keck data points at 1.64~\ms show the difference in I/F when the
halo is included or not; the I/F for all other data points includes both
contributions. The Keck data point near 3.8~\ms is a 1-$\sigma$ upper limit.
(Adapted from \cite{wong06}). 
}
\label{fig:wongspec}}
\end{figure}

Spectra of the rings and variations in intensity with phase angle
provide information on the composition and size distribution of the ring
particles, both of which are tied to the source of the rings. Observations by a
variety of spacecraft and Earth-based telescopes at backscatter all show the
ring to be very red from 0.4 -- 4~\ms (Fig.~\ref{fig:wongspec}).

A variety of independent studies concluded that the color is indicative of the
parent bodies' intrinsic color, and not an effect of scattering by small dust
grains \citep{depater99, throop04, wong06}. The parent bodies may be as red as
Metis, a satellite that is $\sim$~3 times brighter at 2.1~\ms than at visual
wavelengths, \citep{depater99}. The 3.8~\ms 1-$\sigma$ upper limit in
Fig.~\ref{fig:wongspec} is consistent with a deep absorption feature that is
also seen in a spectrum of Amalthea \citep{wong06}.

Images at 1.64 and 2.27~\ms further reveal that the halo contribution is
stronger at the shorter wavelengths, possibly due to an intrinsic red color of
the ring's parent bodies compared with the dusty halo component. 

\begin{figure}
{\figurebox{3.2in}{}{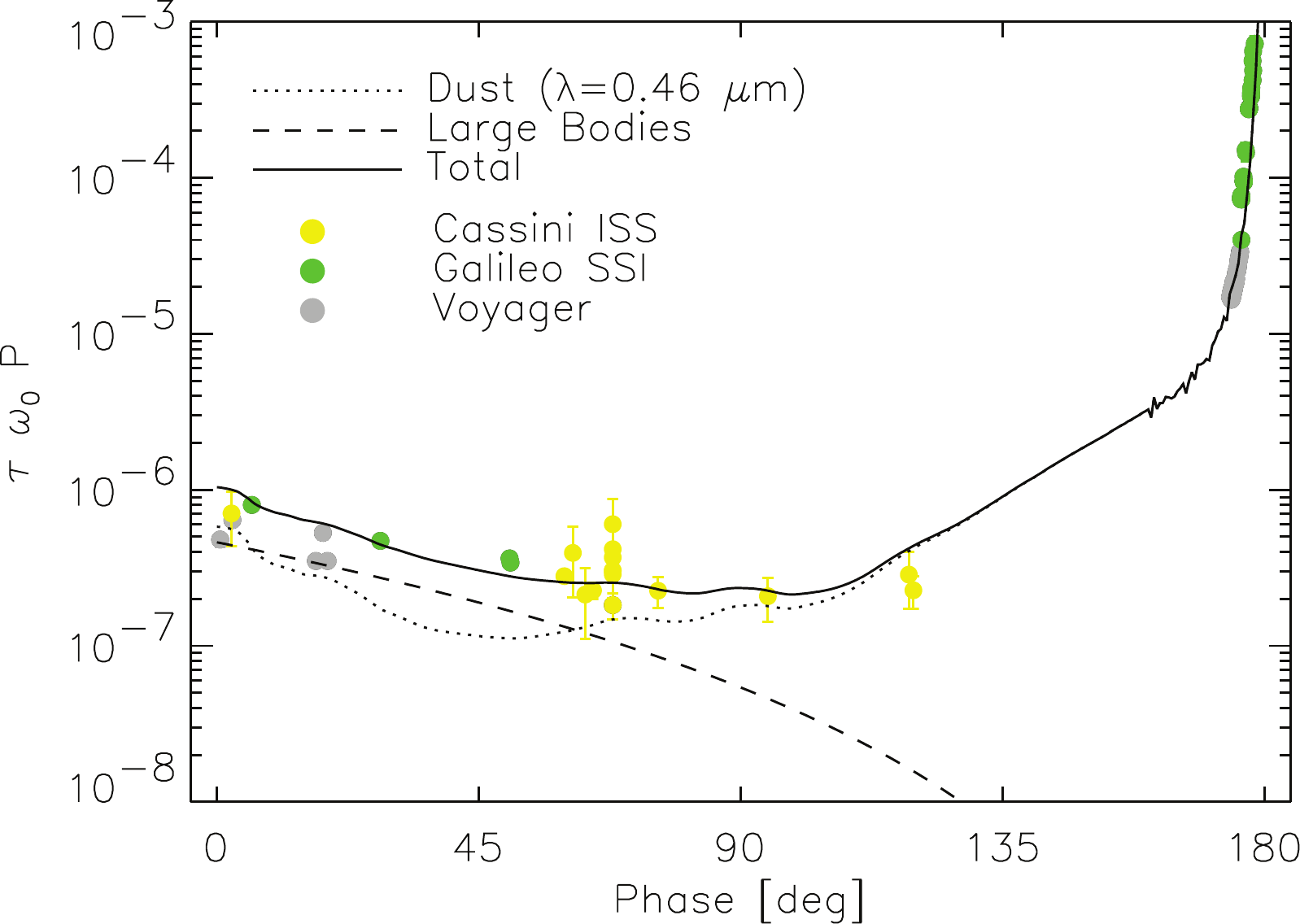}
\caption{Cassini, Galileo and Voyager data as a function of phase angle, with
superposed curves from Throop et al.'s (2004) best-fit non-spherical particle
model at a wavelength of 0.46~\m: the dotted line represents the dust
component, the dashed line the large-body component, and the solid line the sum
of the two. The small ripples in the dust phase curve are due to the finite
number of oblateness values used in the model. \citep{throop04}
}
\label{fig:throop}}
\end{figure}

At high phase angles ($\alpha > 120^\circ$), the observed spectrum is dominated
not by the intrinsic color of a body, but by the scattering properties of small
dust grains.  \cite{throop04} modeled the optical properties of dust
to fit a merged dataset including observations from Galileo, Cassini, HST, and
ground-based telescopes, over phase angles $\alpha = 1^\circ - 179^\circ$
(Fig.~\ref{fig:throop}). They modeled both the spectra and phase angle
dependence with a power law differential particle size distribution:
\begin{equation}
N(r) dr  = N_0 r^{-q} dr
\end{equation}

\noindent with $q \approx 2$ for particles with radii $r<$ 15~\m,
steepening to $q \approx 5$ for larger particles, with a maximum
number density at $r \approx 15\ \mu$m. The red color of the rings,
however, could not be explained by the ring's dust population alone, but
required the distinctly red color of parent-sized material alluded to
above.  \cite{throop04} concluded that the main ring is composed of a
combination of small grains with a normal optical depth
$\tau \approx 4.7 \times 10^{-6}$, and larger bodies at
$\tau \approx 1.3 \times 10^{-6}$. Since the phase curve is rather
flat between 1$^\circ$ and $\sim 130^\circ$, the authors concluded
that the dust grains must be irregular-sized rather than spherical.

Most of the opacity due to large-sized material ($\tau \approx 1.1 \times
10^{-6}$) is contributed by a collection of parent bodies with radii over $\sim
5$ cm, according to a study by \cite{depater08}, who in addition to Keck and
Galileo data also used model fits of Jupiter's synchrotron radiation to
microwave observations. The latter authors further suggested that this
constitutes a fraction of $\sim$~15\% of the total optical depth in the main
ring annulus. The additional optical depth is provided by dust grains
tens--hundreds of $\mu$m across ($\tau \sim 3 \times 10^{-6}$), and another
$\tau \sim 4 \times 10^{-6}$ is attributed to the small (sub)micron-sized
grains that migrate radially inwards due to Poynting-Robertson drag. These
small dust grains form the inward extension of the ring and the halo. 

\subsection{Models of the Jovian Ring System}

Two different models have been proposed for the formation of the jovian ring
system. The first is the model proposed by \cite{burns99}, which was briefly
discussed in the introduction. In this model, dust is knocked off the various
moons in the system, and subsequently evolves inwards under Poynting-Robertson
drag.  The orbit inclination and eccentricity are conserved in this process, and the particles
essentially stay on their near-circular orbits. In the jovian plasma environment, small grains get charged
and their orbital inclinations increase when they cross the 3:2 Lorentz
resonance, producing the halo. 
In an alternative model,
\cite{horanyi96} proposed that the dust's inward migration is caused by plasma
drag, which should lead to substantial changes in the eccentricity of a
particle's orbit. The large eccentricities would result in a concentration of
particles in a thin disk that extends radially from $\sim 80,000$ km out to
$\sim 140,000$ km. However, although observations do suggest the
existence of a thin disk interior to the main ring, none of the observations
show evidence for a thin disk that extends beyond the orbit of Adrastea.

\subsection{Impacts and Time Variability}

Recent studies have shown that the rings of Jupiter and Saturn can
preserve a record of recent impacts \citep{hedman11, showalter11}.
Subtle but distinctive spiral patterns can be
traced back to the moment when the cloud of dust associated with a
passing comet collided with the ring particles. \cite{hedman07}
were the first to identify a pattern produced by a ring impact event.
Cassini images of Saturn's innermost D ring revealed a clear pattern of
periodic, alternating bright and dark bands determined to be vertical
corrugations or ``ripples''. The wavelength of this pattern decreased over
a timeframe of years, due to the differential regression rate of particles on orbits
that initially shared a common node. Because the nodal rate is a known
function of Saturn's gravity moments, the impact date could be pinpointed
to late 1983.

\begin{figure}
{\figurebox{3.2in}{}{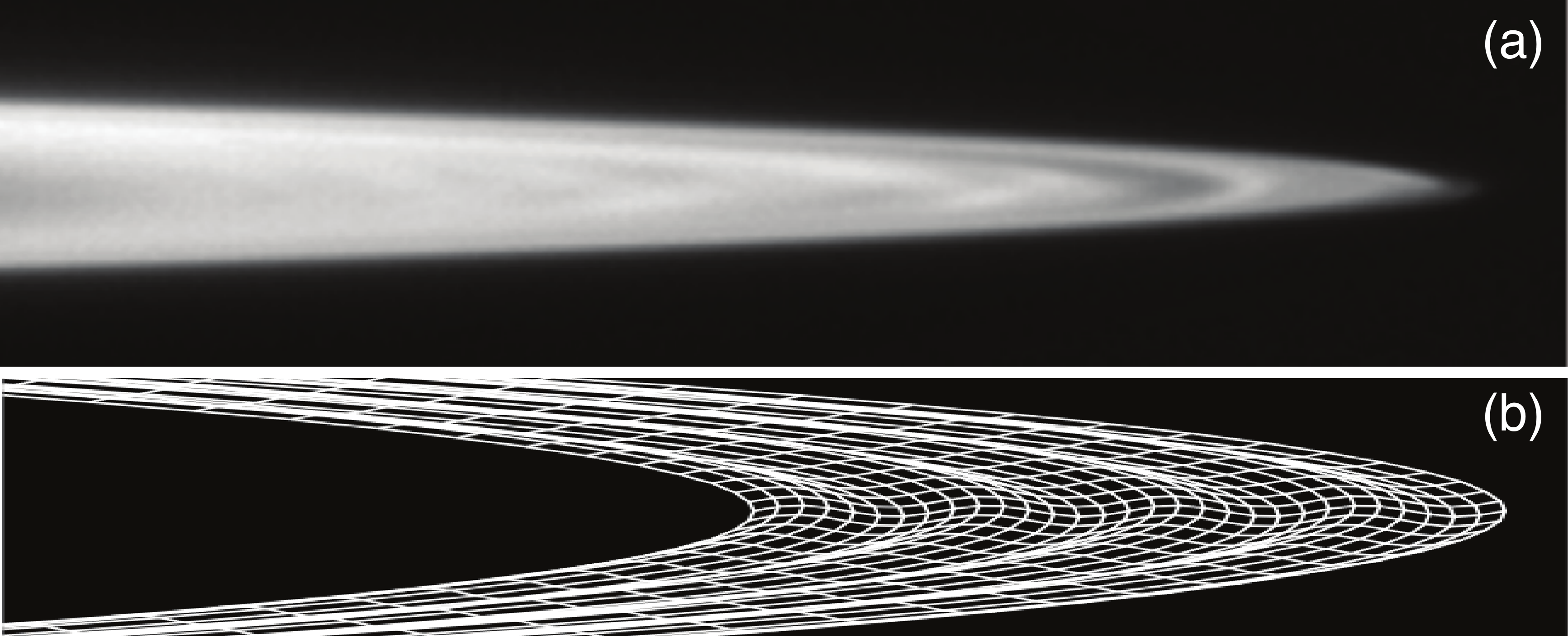}
  \caption{(a) After substantial contrast enhancement, Galileo image C0368974139 from November 1996 reveals subtle reversals of contrast across the ring ansa.
(b) A “cartoon” illustration of how these reversals arise in nearly edge-on views of a corrugated ring. (see also Fig. 1 of \cite{showalter11})
}
\label{fig:ripple}}
\end{figure}


\cite{showalter11} showed that the Galileo spacecraft imaged a
similar pattern of ripples in the jovian ring (Fig.~\ref{fig:ripple}). In this case
they found a superposition of two patterns with different wavelengths,
1900 km and 600 km. The stronger of these patterns had a vertical
amplitude of $\sim$2.4 km and originated in mid-1994. Showalter \et
associated this ``ring-tilting'' event with the Shoemaker-Levy 9 (SL9)
impacts in late July of that year. The secondary pattern originated four
years earlier, and might have been caused by an earlier passage of SL9
through the jovian system. Furthermore, analysis of New Horizons images
from 2007 revealed a pair of similar, but newer, ripple patterns in the
ring, probably caused by smaller and previously unknown ring impacts
during 2001 and 2003.

As the Solar System's largest planet, Jupiter receives sizable cometary
impacts that are now being
observed on a regular basis \citep{hueso13}. History also records four comets
other than SL9 that have had very close brushes with Jupiter;
in 1886, comet P/Brooks 2 may well have scored a direct hit
on Jupiter's main ring. \cite{karydones96} showed that many
comets are temporarily captured into a Jupiter orbit but never strike the
planet. Any of these impacts and near-misses could
potentially tilt the rings.
Based on the four ripple patterns now observed, the time scale
of cometary impacts producing observable ring displacements can be
estimated at 1--2 per decade. Although the scars from planetary impacts
dissipate on a time scale of weeks, each impact into the ring can
produce a spiral pattern potentially observable for 5--10 years. As
a result, the jovian ring may be the most effective ``comet detector''
in the Solar System.

These events may shed light on a long history of suggested, but often
unconfirmed, variabilities in the jovian ring. For example,
Voyager and Galileo clearly imaged ``quadrant asymmetries'' in the main ring
\citep{showalter87, ockert-bell99}, but Cassini and New
Horizons did not \citep{throop04}.
Cassini images showed a 1000-km clump orbiting within the ring \citep{throop04},
unlike anything seen before or since.
During the New Horizons flyby, two clusters of very tiny clumps of unknown
origin were seen within the main ring \citep{showalter07}, but these
have not been seen in other data sets. Jupiter's high cometary influx may
provide part of the explanation for why the jovian ring shows such variability.

\section{The Gossamer Rings}

\begin{figure}
{\figurebox{3.2in}{}{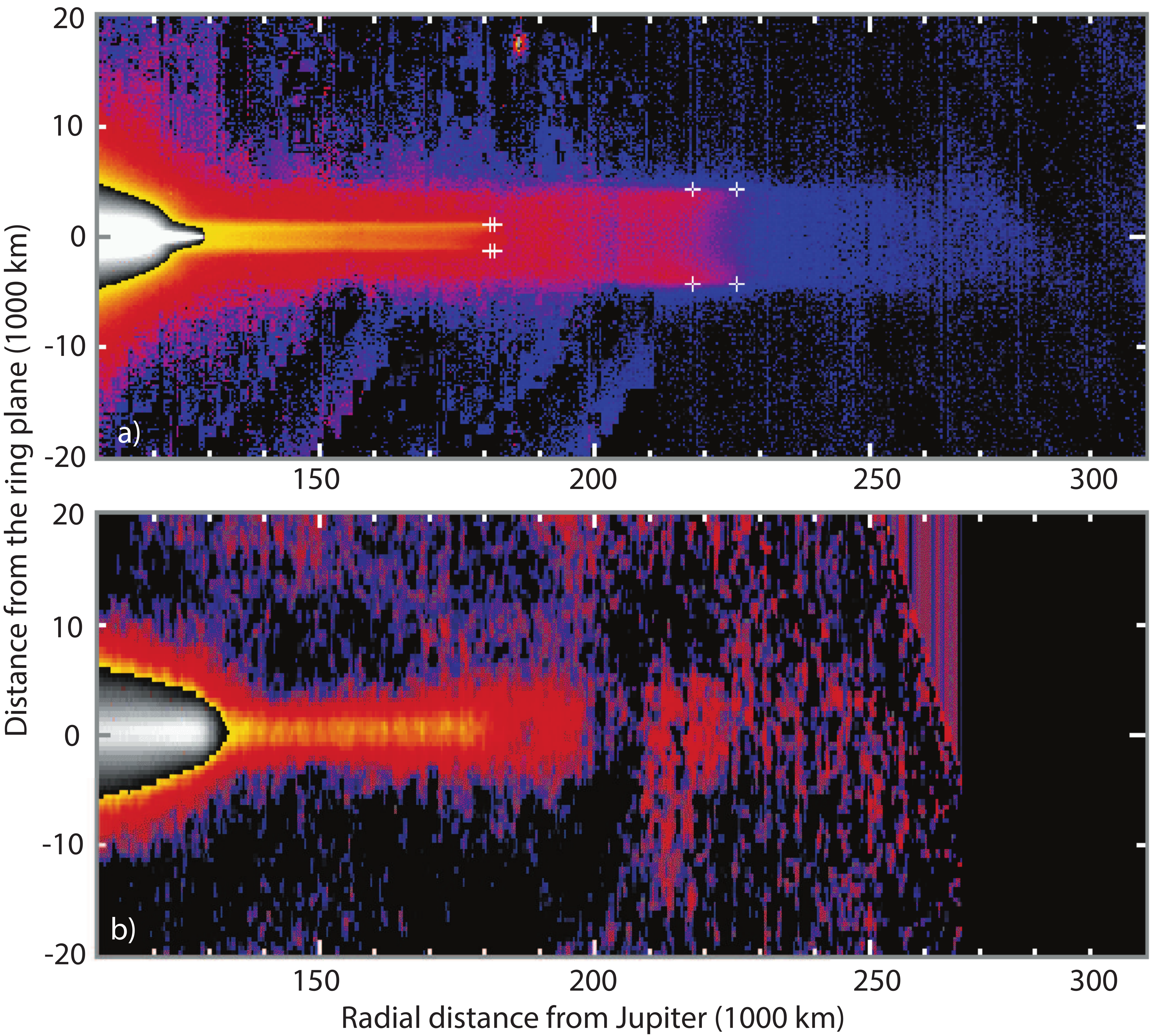}
  \caption{a) Mosaic of Jupiter's gossamer ring system obtained with the
  Galileo spacecraft at high phase angles, 177--179$^\circ$ (mosaic C10).
  Jupiter lies to the left with the main ring and halo in white, the Amalthea
  ring in yellow, the Thebe ring in red, and the Thebe extension in blue. The
  logarithm of the brightness is shown to reduce the dynamic range.  The top
  and bottom edges of the gossamer rings are twice as bright as their central
  cores.  White crosses mark the limits of Amalthea and Thebe's radial and
  vertical motions. b) Jupiter's gossamer rings at a back-scattering phase
  angle of 1.1$^\circ$ taken at 2.27~\ms with Keck in August 1997, close to the
  time that the Earth crossed Jupiter's ring plane (the ring opening angle
  $B=0.17^\circ$). The geometric and color scale was chosen such as to match
  the Galileo image in panel a. The 0.60$^{\prime\prime}$ seeing corresponds to
  a resolution of 1800 km. The main ring, halo, and both gossamer components
  are apparent; hints of gossamer material are also visible further outward in
  the Keck data, albeit barely above the noise level, to the frame's edge at
  $\sim$~257,000 km.  \citep{burns99}}
\label{fig:ring-gos}}
\end{figure}

\begin{figure}
{\figurebox{3.2in}{}{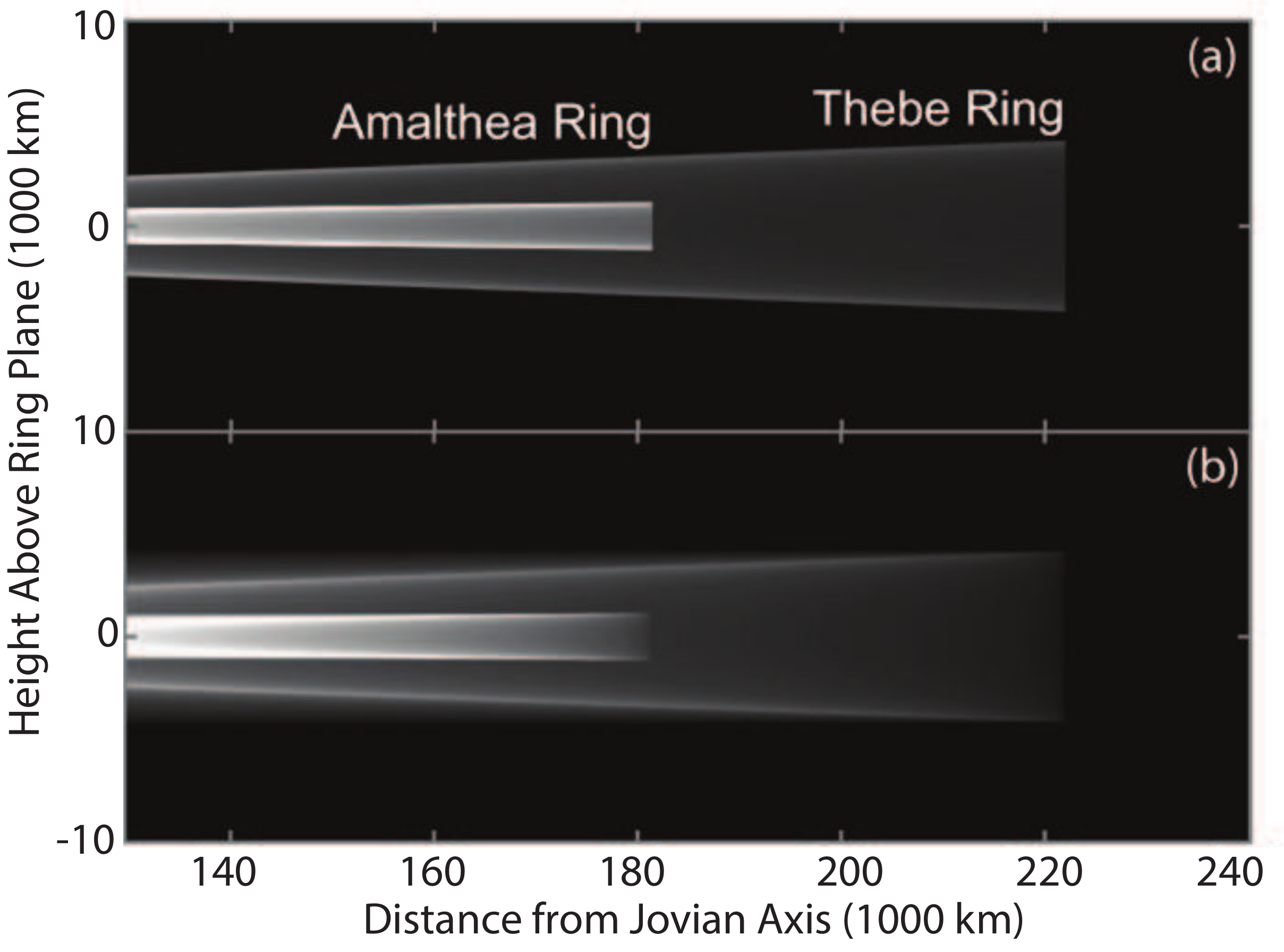}
  \caption{Model images of the gossamer rings. The overlapping
    Amalthea and Thebe rings are shown in (a) cross-sectional and (b)
    edge-on views. The density functions $h(r, z)$ have
    been scaled so that both rings have the same normal optical
    depths; the Thebe ring is fainter overall here only due to its
    greater vertical thickness. The images have been expanded
    vertically by a factor of two to better show the rings’ predicted
    vertical structure. \citep{showalter08} }
\label{fig:show-mod}}
\end{figure}

\begin{figure}
{\figurebox{3.2in}{}{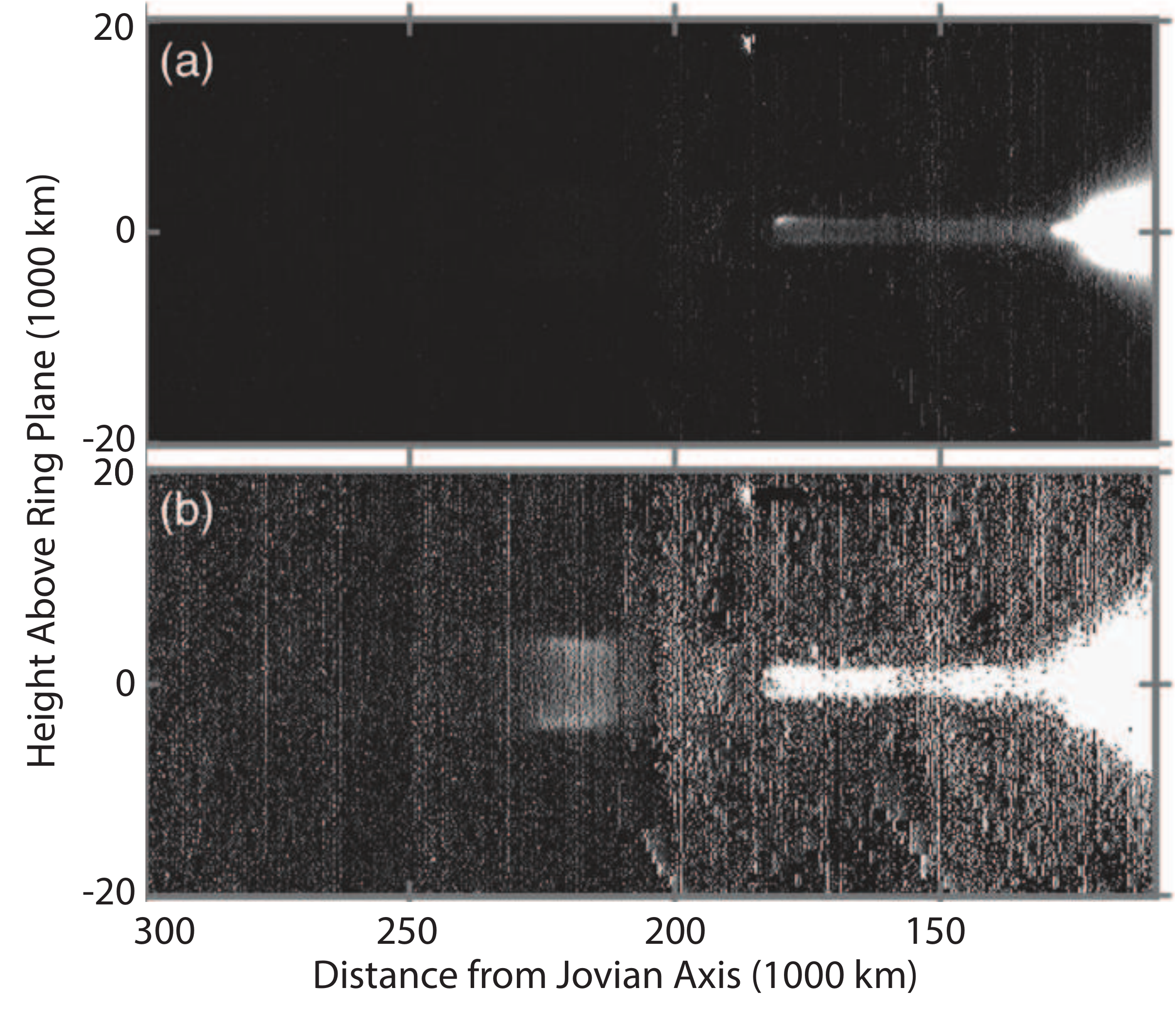}
\caption{An “onion-peeled” image from Fig.~\ref{fig:ring-gos}a. The same image
is shown with two enhancements, one emphasizing the Amalthea ring (a) and the
other emphasizing the Thebe ring (b). The most surprising feature of Amalthea's
ring is a bright peak in intensity near the northern tip. The Thebe ring may
show an inner edge, but this is uncertain because it falls too close to the
border between two images of the mosaic. \citep{showalter08}
}
\label{fig:show-gos}}
\end{figure}

\begin{figure}
{\figurebox{3.2in}{}{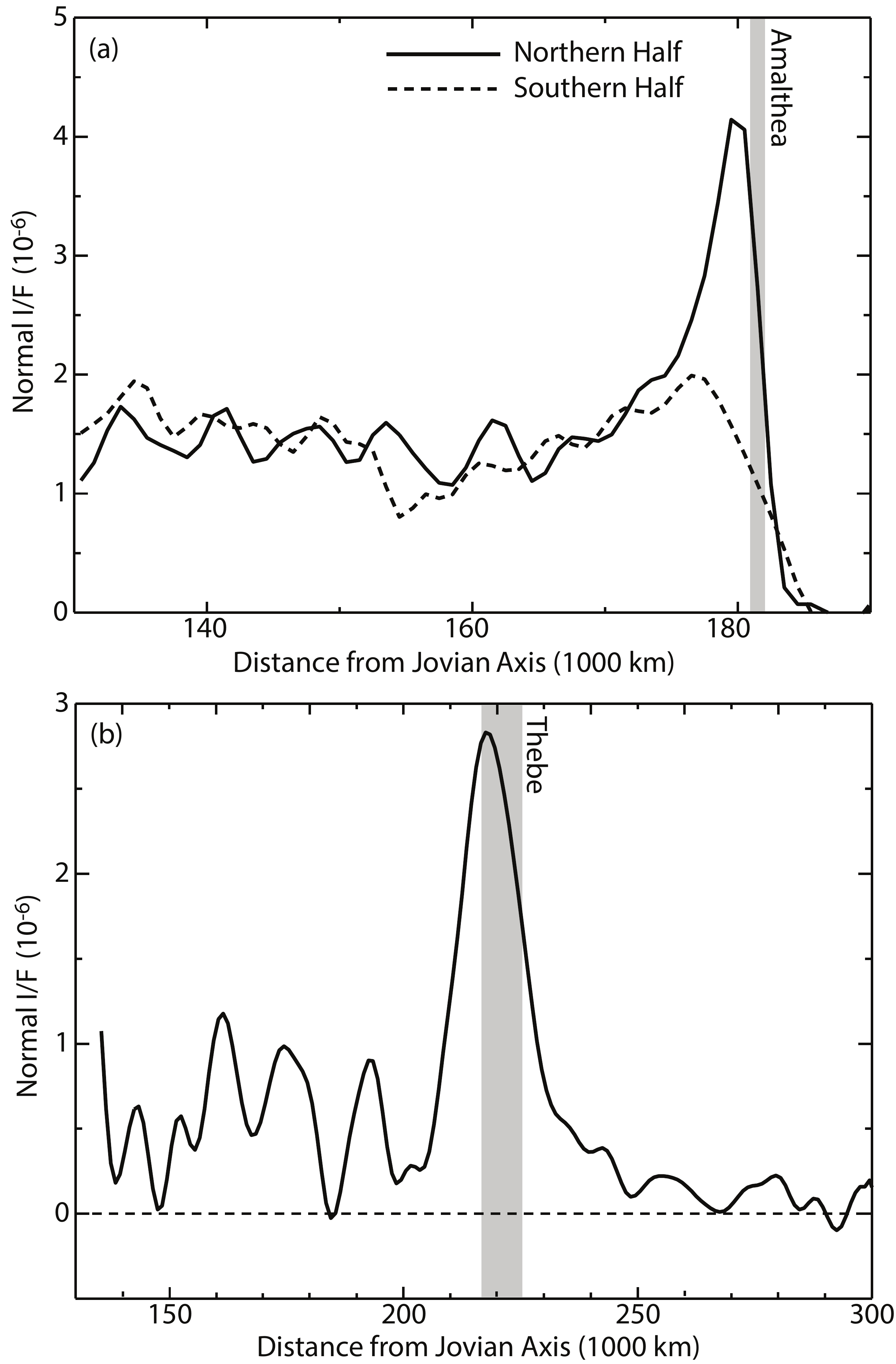}
  \caption{Onion-peeled radial profiles from the Galileo data.  Before
  onion-peeling, to increase the signal-to-noise, an edge-on profile was
  created by vertically integrating rows on the image in
  Fig.~\ref{fig:ring-gos}a. a) Profiles through the northern and southern
  halves of the Amalthea ring: as in Fig.~\ref{fig:show-gos}, the bright peak
  is only shown in the northern half. It is located close but interior to the
  orbit of Amalthea.  b) The Thebe ring: a clear peak is visible just interior
  to Thebe's orbit.  \citep{showalter08}
  }
  \label{fig:show-rad}}
\end{figure}

\begin{figure}
{\figurebox{3.2in}{}{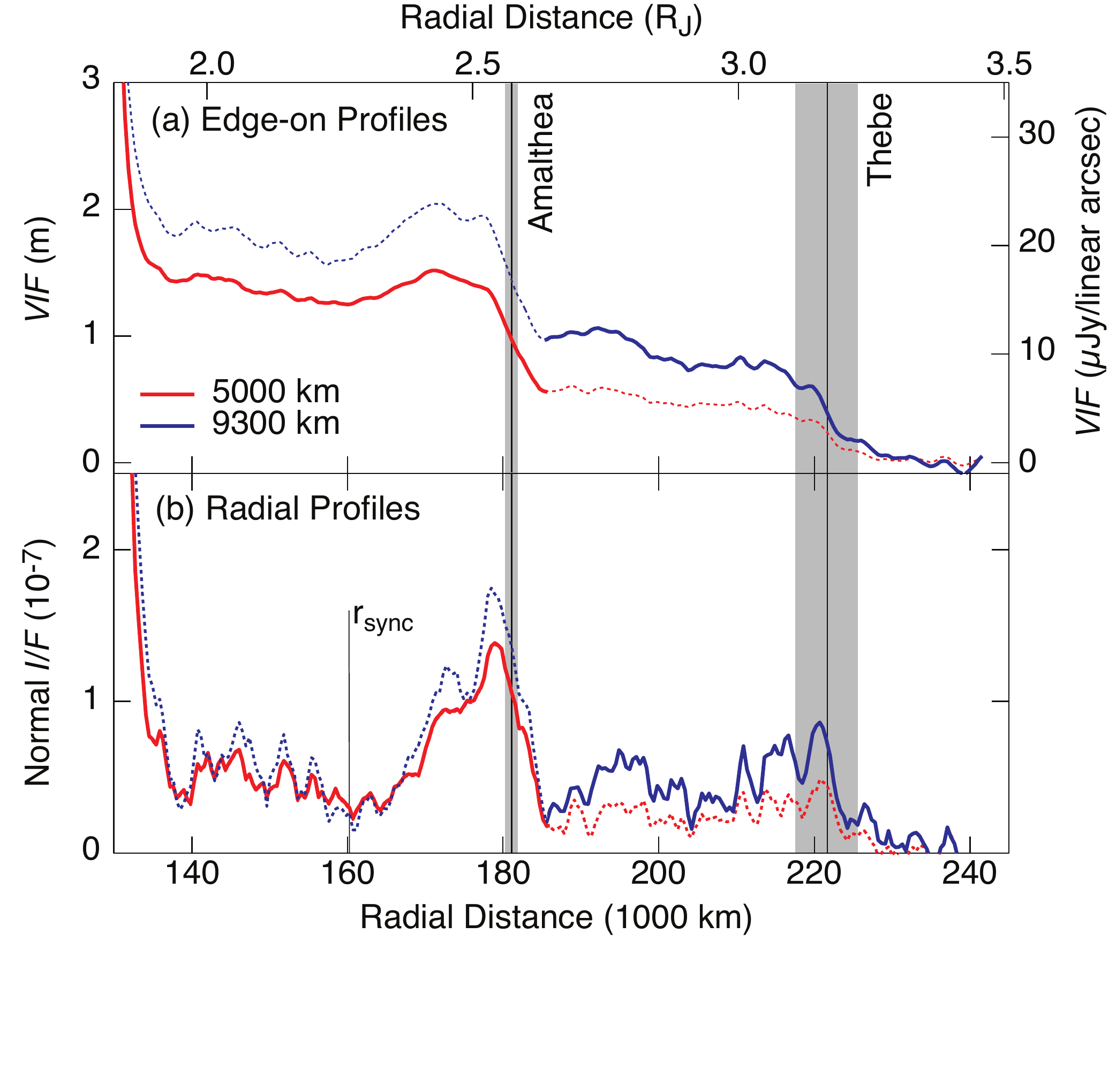}
  \caption{ a) Edge-on profiles through the gossamer rings in Fig.~\ref{fig:keck1}a,
    integrated vertically over $\sim$~1330 km, 5000 km, and 9300 km,
    as indicated. The orbits of Amalthea and Thebe are
    indicated by the vertical grey bars.  b) Onion-peeled radial profiles
    from panel a. The orbits of Amalthea and Thebe, as well as the
    location of synchronous rotation ($r_{\rm sync}$) are
    indicated. The solid vs dashed lines indicate which profiles are
    most appropriate for which ring, as the vertical extent of the
    Thebe ring is about twice as large as for the Amalthea ring
    \citep{depater08}.}
\label{fig:gos-rad}}
\end{figure}

The faint gossamer rings ($\tau \sim 10^{-7}$) consist of two parts
(Fig.~\ref{fig:ring-gos}): the `Amalthea ring' lies immediately interior to the
orbit of Amalthea (at 2.54 $R_\mathrm{J}$); seen edge-on, this ring is almost uniform in
brightness in both forward (panel a) and backscattered (panel b) light.
Interior to Thebe (at 3.11 $R_\mathrm{J}$) is the `Thebe ring', even fainter than the
Amalthea ring. This ring extends inwards and overlaps with the Amalthea ring,
and outwards to well beyond Thebe, out to 3.8 $R_\mathrm{J}$, albeit at a brightness
$\sim$10\% of that of the main Thebe ring. As shown in
Fig.~\ref{fig:ring-gos}a, the upper and lower edges of the gossamer rings are
much brighter than their central cores. The vertical location of the peak
brightness of each of the gossamer rings, as well as their vertical extent,
corresponds to the maxima in the vertically projected distance of the inclined
orbits of Amalthea and Thebe, as indicated by the white crossess on the image.
These characteristics imply that the particles originate from the bounding
satellites, as postulated by the \cite{burns99} model.

\cite{showalter08} re-analyzed the Voyager and Galileo data, together with HST
and Keck (Fig.~\ref{fig:keck1}) images. The Galileo data were taken in
forward-scattered ($\alpha = 170-178^\circ$) and back-scattered ($\alpha \sim
0$--$11^\circ$) light at visible and near-infrared wavelengths. The Keck and
HST observations were taken in back-scattered light at near-infrared (2.2~\m)
and visible wavelengths, respectively. Just like the main ring and halo
(Fig.~\ref{fig:wongspec}), the gossamer rings are $\sim 3$ times brighter in
the infrared than at visible wavelengths, indicative of a power law $q \approx
2$ in eq. 1. An evaluation of Galileo and Voyager phase curves at high phase
angles also indicates $q \sim$~2--2.5 (Showalter et al., 2008).  The very low
backscatter reflectivity of the ring, and a flat phase curve of the ring at low
phase angles (as for the main rings, Fig.~\ref{fig:throop}), suggests that the
ring be composed of distinctly non-spherical particles \citep{showalter08}.

An onion-peeled image of the Keck data was shown in Fig.~\ref{fig:keck1} panels
d and e, and an onion-peeled image of the Galileo mosaic
(Fig.~\ref{fig:ring-gos}a) is shown in Fig.~\ref{fig:show-gos}. These images
look very different: in the Keck image (backscattered light at 2.3~\m), the
Amalthea ring appears to be visible only near the satellite orbit, while the
Galileo image (forward scattered visible-wavelength light) shows a structure
very similar to that seen edge-on. The most striking feature in the latter
image is the bright streak at the northern tip of Amalthea's ring. Onion-peeled
radial profiles from the Galileo data are shown in Fig.~\ref{fig:show-rad},
while Fig.~\ref{fig:gos-rad} shows edge-on and onion-peeled radial profiles
derived from the Keck edge-on image (Fig.~\ref{fig:keck1}a) by integrating
vertically over different heights.  The Galileo-derived Amalthea ring profiles
show that the bright feature is shifted inward from Amalthea's orbit by
$\sim$~1000 km, and is roughly $\sim$~10,000 km wide. Considering the viewing
geometry of the spacecraft and the actual position of Amalthea at the time of
the observations, the detection of this material only along the northern edge
was interpreted as being caused by a population of material trapped in
Lagrangian points along Amalthea's orbit (Showalter \et 2008).
Figure~\ref{fig:show-rad}b reveals a similar peak in Thebe's ring, also shifted
inwards, but symmetric about the equator. Given Galileo's viewing geometry,
Showalter \et (2008) show that this peak can also be explained by material in a
1:1 resonance with Thebe.

The Keck-derived radial profiles are not that dissimilar from the Galileo
profiles, in that a clear peak in brightness is seen just interior of
Amalthea's orbit, although somewhat stronger (relative to the background
profile) and broader than in the Galileo profiles. However, these data were
integrated over 8--9 hours on different nights \citep{depater08}, which could
possibly explain such differences. Moreover, as these data are taken in
backscattered light, they are more sensitive to larger-sized material than the
few up to $\sim$~5~\m-sized dust grains sensed by the Galileo orbiter \citep{krueger09}.
Although the Thebe ring is clearly detected in the Keck profiles, the low
signal-to-noise precludes an affirmative statement with regard to a potential
concentration of particles near Thebe's orbit.  If the minimum in
brightness near the synchronous orbit location (160,250 km) is real, 
plasma drag may play a role in the physics of the particles' orbital
migration.

Aside from these brightness enhancements, the overall profile of the gossamer
rings in the Galileo data is quite uniform, supporting the model of dust
evolving inward at a fixed rate under Poynting-Robertson drag.
Figure~\ref{fig:show-mod} shows model images of this model, which indeed
superficially matches the Galileo data quite well.  Each ring is bounded at its
outer edge by the orbit of its source moon, and has a vertical thickness that
closely matches the thickness as expected for an inclined ring as projected on
the sky.  Also, each ring shows the expected concentrations of material near
its upper and lower edges, and its vertical extent decreases with decreasing
distance to the planet, in agreement with the data.

Although the Amalthea ring appears to be matched very well with the model, the
Thebe ring's vertical profile is less compatible with the model, as the
northern and southern edges are less sharp in the inner part of the Thebe ring
than in the outer part, while both edges are relatively less bright than in the
model. This suggests that the distribution of orbital inclinations of the
particles appears to broaden while the grains move inward; they may receive
kicks when they cross the strong 3:2 and 4:3 exterior Lorentz resonances (at
209,500 and 193,400 km, resp.) on their journey inwards.

\section{In-situ Detection of Dust}

Near the end of its mission, the Galileo orbiter passed through the                                           
Thebe and Amalthea gossamer rings on November 5, 2002 and again on                                            
September 21, 2003. Impacts of numerous particles between 0.2 and 5~\ms 
in radius were detected, extending the size distribution                                              
downward to particles not apparent in visual                                                                  
imaging \citep{krueger09}. In addition to sensitivity to smaller                                               
particles, in-situ impact detections are also more sensitive to lower                                         
absolute number densities than direct imaging. Thus one of the main                                           
results of Galileo's passage through the ring region was the discovery                                        
that dust in the outer Thebe ring extends well beyond the limits of                                           
imaging to at least 5 Jupiter radii. The outer edge of the Thebe ring,                                        
therefore, is likely set by destabilizing gravitational perturbations                                         
from Io, located at 5.9~$R_\mathrm{J}$.                                                                        

Galileo also showed that the power-law size distribution measured                                             
optically in the main rings holds in the gossamer rings, at                                                   
least for the micron-sized particles. Submicron particles follow                                              
this size distribution in the Thebe ring as well, but there is a notable excess                               
of 0.2~\ms particles measured in the Amalthea                                                              
ring \citep{krueger09,hamilton08}.  Finally, the incoming angles of                                            
impacting particles imply that orbital inclination angles for some                                            
grains extend up to at least 20$^\circ$, far exceeding the 1$^\circ$                                           
tilt of large optically visible Thebe grains \citep{krueger09,hamilton08}.                                     

These interesting observations can be explained as due to                                                     
electromagnetic perturbations arising on orbiting dust                                                        
grains \citep{hamilton08}. Dust grains are electrically charged by                                             
interactions with sunlight and local plasma, so their motions are                                             
influenced by Jupiter's strong magnetic field \citep{burns01, burns04}. In particular,                                           
the change in a dust grain's electric charge during passage through                                           
Jupiter's shadow is especially important. As these changes occur once                                         
per orbit, the effect is resonant and is known as a shadow                                                    
resonance. In its original formulation \citep{horanyi1991} the resonance                                        
excites orbital eccentricities, but there is also an analogous                                                
vertical shadow resonance that excites orbital                                                               
inclinations \citep{hamilton08}. The eccentricity resonance can account                                        
for the outward extension of the Thebe ring seen both in imaging and                                          
in-situ while the narrower inclination resonance excites micron-sized                                         
grains to the high tilts implied by Galileo's in-situ                                                         
detections \citep{hamilton08}.  The excess of submicron grains near                                            
Amalthea is likely due to its proximity to Jupiter's synchronous orbit                                           
where electromagnetic perturbations dramatically weaken. More details 
can be found in \citet{hamilton08} and in Chapter~\ref{HedmanDusty} of this book.

\section{Conclusions and Outlook}

The jovian ring has now been observed closely by seven spacecraft, starting
with Pioneer 11's dust detector and including, very recently, Juno. With the inclusion of
Earth-based data from HST and a number of
large ground-based telescopes, the jovian ring system
has been observed repeatedly and in some detail. From these disparate data sets
a reasonably coherent picture of the rings has started to emerge. The rings are composed mostly of
dust that is derived from meteorite impacts on the four ring-moons Thebe,
Amalthea, Metis and Adrastea, as well as via impacts on and collisions between
cm- and larger-sized material trapped in the main ring annulus and perhaps in
1:1 resonances with Amalthea and Thebe. The dust migrates inward due to
Poynting-Robertson drag, as first suggested and modeled by \cite{burns99};
their model seems to explain most of the ring's structure.  Inward-migrating
dust grains are ``kicked'' into a halo at the 2:3 Lorentz resonance; further
inward, at the 1:2 Lorentz resonance, the inclinations of the particles are
increased even more, and grains are ultimately lost into the atmosphere of
Jupiter. Lorentz resonances in the Thebe ring are likely responsible for
discrepancies between the \cite{burns99} model and observations of the Thebe ring.
For example, the outward extension to the Thebe ring may be caused by small grains in
Thebe's ring that receive large eccentricities due to a shadow resonance.


Nevertheless, observations of the ring remain few in number and are quite
sparse in time coverage. Each observing period involves different viewing
geometry and different instruments, often sensitive to different wavelengths.
Most importantly, it involves a different epoch. Meanwhile, numerous ring
features, from ripple patterns to clumps and
asymmetries of various sorts, are time-variable. These phenomena are not
possible to understand fully within such an episodic data set.

When the rings turn edge-on again in March 2021, multi-wavelength observations
with the James Webb Space Telescope (JWST) may help solve several still
outstanding questions about the jovian system. However, we will probably not
obtain a comprehensive understanding of this fascinating system until it can
be scrutinized regularly and in detail by a Jupiter-orbiting spacecraft.

\section*{Acknowledgements}
I.d.P. acknowledges partial support from NASA Planetary Astronomy (PAST) award NNX14AJ43G.
M.R.S. acknowledges support from NASA?s Outer Planets Research Program through grant NNX14AO40G, and from Space Telescope Science Institute through program HST-GO-13414. Support for Program number HST-GO-13414 was provided by NASA through a grant from the Space Telescope Science Institute, which is operated by the Association of Universities for Research in Astronomy, Incorporated, under NASA contract NAS5-26555.
D.P.H. acknowledges support from the Cassini Data Analysis Program Grant NNX15AQ67G.

\bibliography{jupring} 
\bibliographystyle{cambridgeauthordate}

\newcounter{chapterdummy}
\newenvironment{chapterdummy}{\refstepcounter{chapterdummy}}

\textcolor{white}{
\chapterdummy{Space age studies of planetary rings \label{Esposito}}
\chapterdummy{An introduction to planetary ring dynamics \label{HedmanIntro}}
\chapterdummy{The rings of Saturn \label{Saturn}}
\chapterdummy{The rings of Uranus \label{Uranus}}
\chapterdummy{The rings of Neptune \label{Neptune}}
\chapterdummy{The rings of Jupiter \label{Jupiter}}
\chapterdummy{Rings beyond the giant planets \label{Sicardy}}
\chapterdummy{Moonlets in dense planetary rings \label{Spahn}}
\chapterdummy{Meteoroid bombardment and ballistic transport in planetary rings \label{Estrada}}
\chapterdummy{Theory of narrow rings and sharp edges \label{Longaretti}}
\chapterdummy{Narrow rings, gaps, and sharp edges \label{NicholsonEdges}}
\chapterdummy{Dusty rings \label{HedmanDusty}}
\chapterdummy{The F ring of Saturn \label{Murray}}
\chapterdummy{Plasma, neutral atmosphere, and energetic radiation environments of planetary rings \label{Cooper}}
\chapterdummy{Thermal properties of rings and ring particles \label{SpilkerThermal}}
\chapterdummy{Computer simulations of planetary rings \label{Salo}}
\chapterdummy{Laboratory studies of planetary ring systems \label{Colwell}}
\chapterdummy{The origin of planetary ring systems \label{Charnoz}}
\chapterdummy{Future missions to planetary rings \label{SpilkerMissions}}
\chapterdummy{Planetary rings and other astrophysical disks \label{Latter}}
\chapterdummy{The future of planetary rings studies \label{Tiscareno}}}

\end{document}

%% file: Table1.tex

\begin{table*} 
\caption{Properties of Jupiter's ring system.$^{a}$}%
\begin{tabular*}{170mm}{@{}l*{6}{@{\extracolsep{0pt plus 12pt}}l}@{}}
\toprule
& Halo$^b$ & Main   ring & Amalthea  ring & Thebe  ring & Thebe extension \\
\hline
Radial location (R$_J$) & 1.4--1.71 & 1.72--1.806 & 1.8--2.55 & 1.8--3.10 & 3.1--3.8 \\
Radial location (km) & 100\,000--122\,400 & 122\,400--129\,100 & 122\,400--181\,350 & 122\,400--221\,900 & 221\,900--270\,000 \\
Vertical  thickness & ${\sim}5 \times 10^4$  km & 30--100  km & ${\sim}2300$  km & ${\sim}8500$   km & $\sim$9000~km \\
Normal  optical  depth & few $\times$ $10^{-6}$ & few $\times$ $10^{-6}$ & ${\sim}10^{-7}$ & ${\sim}10^{-8}$ & ${\sim}10^{-9}$ \\
Particle  size & (sub)$\mu$m & broad distribution & broad distribution &  broad distribution \\
\botrule
\end{tabular*}
\begin{tabnote}
{$^a$ Table from \cite{depatlis}, based on data from \cite{ockert-bell99}, and\cite{depater99, depater08}. \\[2pt]
$^b$ Numbers quoted are based upon the Galileo data (visible light data,
in forward scattered light). Relative to the main ring, the halo is
much less bright and more spatially confined at longer wavelengths and
in backscattered light.}
\end{tabnote}
\label{tbl:tab1}
\end{table*}

%% file: jupring.bbl
\begin{thebibliography}{29}
\expandafter\ifx\csname natexlab\endcsname\relax\def\natexlab#1{#1}\fi
\expandafter\ifx\csname selectlanguage\endcsname\relax
  \def\selectlanguage#1{\relax}\fi

\bibitem[\protect\citename{{Burns} {et~al.}, }1984]{burns84}
{Burns}, J.~A., {Showalter}, M.~R., and {Morfill}, G.~E. 1984.
\newblock {The ethereal rings of Jupiter and Saturn}.
\newblock {Pages  200--272 of:} {Greenberg}, R., and {Brahic}, A. (eds), {\em
  IAU Colloq. 75: Planetary Rings}.

\bibitem[\protect\citename{{Burns} {et~al.}, }1999]{burns99}
{Burns}, J.~A., {Showalter}, M.~R., {Hamilton}, D.~P., {Nicholson}, P.~D., {de
  Pater}, I., {Ockert-Bell}, M.~E., and {Thomas}, P.~C. 1999.
\newblock {The Formation of Jupiter's Faint Rings}.
\newblock {\em Science}, {\bf 284}(May), 1146.

\bibitem[\protect\citename{Burns {et~al.}, }2001]{burns01}
Burns, J.~A., {Hamilton}, D.~P., and {Showalter}, M.~R. 2001.
\newblock {\em Interplanetary Dust}.
\newblock Springer.
\newblock Chap. Planetary Rings, pages  387--438.

\bibitem[\protect\citename{{Burns} {et~al.}, }2004]{burns04}
{Burns}, J.~A., {Simonelli}, D.~P., {Showalter}, M.~R., {Hamilton}, D.~P.,
  {Porco}, C.~D., {Throop}, H., and {Esposito}, L.~W. 2004.
\newblock {\em {Jupiter's ring-moon system}}.
\newblock Cambridge, UK: Cambridge University Press.
\newblock Pages  241--262.

\bibitem[\protect\citename{{de Pater} and {Lissauer}, }2015]{depatlis}
{de Pater}, I., and {Lissauer}, J.~J. 2015.
\newblock {\em {Planetary Sciences}}.
\newblock Cambridge, UK: Cambridge University Press.

\bibitem[\protect\citename{{de Pater} {et~al.}, }1999]{depater99}
{de Pater}, I., {Showalter}, M.~R., {Burns}, J.~A., {Nicholson}, P.~D., {Liu},
  M.~C., {Hamilton}, D.~P., and {Graham}, J.~R. 1999.
\newblock {Keck Infrared Observations of Jupiter's Ring System near Earth's
  1997 Ring Plane Crossing}.
\newblock {\em \icarus}, {\bf 138}(Apr.), 214--223.

\bibitem[\protect\citename{{de Pater} {et~al.}, }2008]{depater08}
{de Pater}, I., {Showalter}, M.~R., and {Macintosh}, B. 2008.
\newblock {Keck observations of the 2002 2003 jovian ring plane crossing}.
\newblock {\em \icarus}, {\bf 195}(May), 348--360.

\bibitem[\protect\citename{{Fillius} {et~al.}, }1975]{fillius75}
{Fillius}, R.~W., {McIlwain}, C.~E., and {Mogro-Campero}, A. 1975.
\newblock {Radiation belts of Jupiter - A second look}.
\newblock {\em Science}, {\bf 188}(May), 465--467.

\bibitem[\protect\citename{{Hamilton} and {Kr{\"u}ger}, }2008]{hamilton08}
{Hamilton}, D.~P., and {Kr{\"u}ger}, H. 2008.
\newblock {The sculpting of Jupiter's gossamer rings by its shadow}.
\newblock {\em Nature}, {\bf 453}(May), 72--75.

\bibitem[\protect\citename{{Hedman} {et~al.}, }2007]{hedman07}
{Hedman}, M.~M., {Burns}, J.~A., {Showalter}, M.~R., {Porco}, C.~C.,
  {Nicholson}, P.~D., {Bosh}, A.~S., {Tiscareno}, M.~S., {Brown}, R.~H.,
  {Buratti}, B.~J., {Baines}, K.~H., and {Clark}, R. 2007.
\newblock {Saturn's dynamic D ring}.
\newblock {\em \icarus}, {\bf 188}(May), 89--107.

\bibitem[\protect\citename{{Hedman} {et~al.}, }2011]{hedman11}
{Hedman}, M.~M., {Burns}, J.~A., {Evans}, M.~W., {Tiscareno}, M.~S., and
  {Porco}, C.~C. 2011.
\newblock {Saturn{'}s Curiously Corrugated C Ring}.
\newblock {\em Science}, {\bf 332}(May), 708.

\bibitem[\protect\citename{{Horanyi} and {Burns}, }1991]{horanyi1991}
{Horanyi}, M., and {Burns}, J.~A. 1991.
\newblock {Charged dust dynamics - Orbital resonance due to planetary shadows}.
\newblock {\em \jgr}, {\bf 96}(Nov.), 19.

\bibitem[\protect\citename{{Hor{\'a}nyi} and {Cravens}, }1996]{horanyi96}
{Hor{\'a}nyi}, M., and {Cravens}, T.~E. 1996.
\newblock {The structure and dynamics of Jupiter's ring}.
\newblock {\em \nat}, {\bf 381}(May), 293--295.

\bibitem[\protect\citename{{Hueso} {et~al.}, }2013]{hueso13}
{Hueso}, R., {P{\'e}rez-Hoyos}, S., {S{\'a}nchez-Lavega}, A., {Wesley}, A.,
  {Hall}, G., {Go}, C., {Tachikawa}, M., {Aoki}, K., {Ichimaru}, M., {Pond},
  J.~W.~T., {Korycansky}, D.~G., {Palotai}, C., {Chappell}, G., {Rebeli}, N.,
  {Harrington}, J., {Delcroix}, M., {Wong}, M., {de Pater}, I., {Fletcher},
  L.~N., {Hammel}, H., {Orton}, G.~S., {Tabe}, I., {Watanabe}, J., and
  {Moreno}, J.~C. 2013.
\newblock {Impact flux on Jupiter: From superbolides to large-scale
  collisions}.
\newblock {\em \aap}, {\bf 560}(Dec.), A55.

\bibitem[\protect\citename{{Kary} and {Dones}, }1996]{karydones96}
{Kary}, D.~M., and {Dones}, L. 1996.
\newblock {Capture Statistics of Short-Period Comets: Implications for Comet
  D/Shoemaker-Levy 9}.
\newblock {\em \icarus}, {\bf 121}(June), 207--224.

\bibitem[\protect\citename{{Kr{\"u}ger} {et~al.}, }2009]{krueger09}
{Kr{\"u}ger}, H., {Hamilton}, D.~P., {Moissl}, R., and {Gr{\"u}n}, E. 2009.
\newblock {Galileo in-situ dust measurements in Jupiter's gossamer rings}.
\newblock {\em Icarus}, {\bf 203}(Sept.), 198--213.

\bibitem[\protect\citename{{Ockert-Bell} {et~al.}, }1999]{ockert-bell99}
{Ockert-Bell}, M.~E., {Burns}, J.~A., {Daubar}, I.~J., {Thomas}, P.~C.,
  {Veverka}, J., {Belton}, M.~J.~S., and {Klaasen}, K.~P. 1999.
\newblock {The Structure of Jupiter's Ring System as Revealed by the Galileo
  Imaging Experiment}.
\newblock {\em Icarus}, {\bf 138}(Apr.), 188--213.

\bibitem[\protect\citename{Showalter, }2001]{showalter01}
Showalter, M.~R., Hamilton D.~P. Burns J.~A. de Pater I. Simonelli D.~P. 2001.
\newblock Structure of Jupiter's main ring and halo from Galileo SSI and
  Earth-based images.
\newblock {Pages  100--101 of:} {\em Planet, Satellites and Magnetosphere}.

\bibitem[\protect\citename{{Showalter} {et~al.}, }1985]{showalter85}
{Showalter}, M.~R., {Burns}, J.~A., {Cuzzi}, J.~N., and {Pollack}, J.~B. 1985.
\newblock {Discovery of Jupiter's 'gossamer' ring}.
\newblock {\em \nat}, {\bf 316}(Aug.), 526--528.

\bibitem[\protect\citename{{Showalter} {et~al.}, }1987]{showalter87}
{Showalter}, M.~R., {Burns}, J.~A., {Cuzzi}, J.~N., and {Pollack}, J.~B. 1987.
\newblock {Jupiter's ring system - New results on structure and particle
  properties}.
\newblock {\em \icarus}, {\bf 69}(Mar.), 458--498.

\bibitem[\protect\citename{{Showalter} {et~al.}, }2007]{showalter07}
{Showalter}, M.~R., {Cheng}, A.~F., {Weaver}, H.~A., {Stern}, S.~A., {Spencer},
  J.~R., {Throop}, H.~B., {Birath}, E.~M., {Rose}, D., and {Moore}, J.~M. 2007.
\newblock {Clump Detections and Limits on Moons in Jupiter{'}s Ring
  System}.
\newblock {\em Science}, {\bf 318}(Oct.), 232.

\bibitem[\protect\citename{{Showalter} {et~al.}, }2008]{showalter08}
{Showalter}, M.~R., {de Pater}, I., {Verbanac}, G., {Hamilton}, D.~P., and
  {Burns}, J.~A. 2008.
\newblock {Properties and dynamics of Jupiter's gossamer rings from Galileo,
  Voyager, Hubble and Keck images}.
\newblock {\em Icarus}, {\bf 195}(May), 361--377.

\bibitem[\protect\citename{{Showalter} {et~al.}, }2011]{showalter11}
{Showalter}, M.~R., {Hedman}, M.~M., and {Burns}, J.~A. 2011.
\newblock {The Impact of Comet Shoemaker-Levy 9 Sends Ripples Through the Rings
  of Jupiter}.
\newblock {\em Science}, {\bf 332}(May), 711.

\bibitem[\protect\citename{{Smith} {et~al.}, }1979a]{smith79b}
{Smith}, B.~A., {Soderblom}, L.~A., {Beebe}, R., {Boyce}, J., {Briggs}, G.,
  {Carr}, M., {Collins}, S.~A., {Johnson}, T.~V., {Cook}, II, A.~F.,
  {Danielson}, G.~E., and {Morrison}, D. 1979a.
\newblock {The Galilean satellites and Jupiter - Voyager 2 imaging science
  results}.
\newblock {\em Science}, {\bf 206}(Nov.), 927--950.

\bibitem[\protect\citename{{Smith} {et~al.}, }1979b]{smith79a}
{Smith}, B.~A., {Soderblom}, L.~A., {Johnson}, T.~V., {Ingersoll}, A.~P.,
  {Collins}, S.~A., {Shoemaker}, E.~M., {Hunt}, G.~E., {Masursky}, H., {Carr},
  M.~H., {Davies}, M.~E., {Cook}, A.~F., {Boyce}, J.~M., {Owen}, T.,
  {Danielson}, G.~E., {Sagan}, C., {Beebe}, R.~F., {Veverka}, J., {McCauley},
  J.~F., {Strom}, R.~G., {Morrison}, D., {Briggs}, G.~A., and {Suomi}, V.~E.
  1979b.
\newblock {The Jupiter system through the eyes of Voyager 1}.
\newblock {\em Science}, {\bf 204}(June), 951--957.

\bibitem[\protect\citename{{Throop} {et~al.}, }2004]{throop04}
{Throop}, H.~B., {Porco}, C.~C., {West}, R.~A., {Burns}, J.~A., {Showalter},
  M.~R., and {Nicholson}, P.~D. 2004.
\newblock {The jovian rings: new results derived from Cassini, Galileo,
  Voyager, and Earth-based observations}.
\newblock {\em Icarus}, {\bf 172}(Nov.), 59--77.

\bibitem[\protect\citename{Throop {et~al.}, }2016]{throop2016}
Throop, H.~B., Showalter, M.~R., Dones, H.~C., Hamilton, D.~P., Weaver, H.~A.,
  Cheng, A.~F., Stern, S.~A., Young, L., Olkin, C.~B, and {New Horizons Science
  Team}. 2016.
\newblock {New Horizons Imaging of Jupiter's Main Ring}.
\newblock {\em AAS/Division for Planetary Sciences Meeting}, {\bf 48}(Oct.).

\bibitem[\protect\citename{{van Allen} {et~al.}, }1975]{vanallen75}
{van Allen}, J.~A., {Randall}, B.~A., {Baker}, D.~N., {Goertz}, C.~K.,
  {Sentman}, D.~D., {Thomsen}, M.~F., and {Flindt}, H.~R. 1975.
\newblock {Pioneer 11 observations of energetic particles in the Jovian
  magnetosphere}.
\newblock {\em Science}, {\bf 188}(May), 459--462.

\bibitem[\protect\citename{{Wong} {et~al.}, }2006]{wong06}
{Wong}, M.~H., {de Pater}, I., {Showalter}, M.~R., {Roe}, H.~G., {Macintosh},
  B., and {Verbanac}, G. 2006.
\newblock {Ground-based near infrared spectroscopy of Jupiter's ring and
  moons}.
\newblock {\em \icarus}, {\bf 185}(Dec.), 403--415.

\end{thebibliography}
